\documentclass[10pt,a4paper]{article}
\usepackage{amsmath,latexsym,amssymb,amsfonts}
\usepackage[dvips]{graphicx}
\usepackage{bm}

\usepackage{color}

\newcommand{\beq}{\begin{equation}}
\newcommand{\eeq}{\end{equation}}
\newcommand{\bea}{\begin{eqnarray}}
\newcommand{\eea}{\end{eqnarray}}

\newcommand{\p}{{\rm P}}

\begin{document}

\title{\textbf{Toward inflation models compatible with the no-boundary proposal}}
\author{\textsc{Dong-il Hwang}$^{a,b}$\footnote{dongil.j.hwang@gmail.com}\; and \textsc{Dong-han Yeom}$^{a,c}$\footnote{innocent.yeom@gmail.com}\\
\textit{$^{a}$\small{Center for Quantum Spacetime, Sogang University, Seoul 121-742, Republic of Korea}}\\
\textit{$^{b}$\small{Research Institute for Basic Science, Sogang University, Seoul 121-742, Republic of Korea}}\\
\textit{$^{c}$\small{Yukawa Institute for Theoretical Physics, Kyoto University, Kyoto 606-8502, Japan}}
}

\maketitle

\begin{abstract}
In this paper, we investigate various inflation models in the context of the no-boundary proposal. We propose that a good inflation model should satisfy three conditions: observational constraints, plausible initial conditions, and naturalness of the model. For various inflation models, we assign the probability to each initial condition using the no-boundary proposal and define a quantitative standard, typicality, to check whether the model satisfies the observational constraints with probable initial conditions. There are three possible ways to satisfy the typicality criterion: there was pre-inflation near the high energy scale, the potential is finely tuned or the inflationary field space is unbounded, or there are sufficient number of fields that contribute to inflation. The no-boundary proposal rejects some of naive inflation models, explains some of traditional doubts on inflation, and possibly, can have observational consequences.
\end{abstract}
\begin{flushright}
{\tt YITP-13-122}
\end{flushright}

\newpage

\tableofcontents

\newpage
\section{Introduction}

To resolve the traditional problems of big bang cosmology, \textit{inflation} has been the most compelling idea since a few decades ago \cite{Guth:1980zm,Starobinsky:1980te}. Nevertheless, inflation is not free from the problem of initial conditions and we need to explain why and how such inflation occurred \cite{Linde:1981mu}. For this purpose, we need to specify field contents, couplings between each field, parameters of the potential. Although there are observational constraints and recently refined by the Planck satellite mission \cite{Ade:2013uln}, if a model is flexible on parameters or assumptions, then there still remain many models that are possible to explain the observed universe. According to the recent data, a simple form of chaotic inflation \cite{Linde:1983gd} seems not be preferred while a convex potential is rather preferred (e.g., \cite{Starobinsky:1980te}).

However, this is not enough in general. If the initial condition for the model is not natural and rather tricky, then we need a justification or we should regard the hypothesis as unconvincing. Then, can we give a good -- \textit{quantitative} -- standard to check whether a model/hypothesis is good or bad? In principle, this is possible from \textit{quantum cosmology}. In quantum cosmology, we canonically quantize the metric and the matter at the same time and define the wave equation of the universe, through the Wheeler-DeWitt equation \cite{DeWitt:1967yk}. Traditionally, quantum cosmology was developed to explain the initial singularity \cite{Hawking:1969sw}; nowadays, quantum cosmology and the wave function of the universe are useful to assign a probability for a certain set of initial field configurations. Therefore, quantum cosmology is useful to study the initial condition of inflation, as well as to build a qualitative standard to assess a hypothesis.

Now the question is that what is the wave function of our universe. The Wheeler-DeWitt equation is a partial differential equation and hence requires the boundary condition (for general review, see \cite{Kiefer}). The probability distribution will depend on the choice of the boundary condition. Who knows the boundary condition? Although no one can give a definite answer, perhaps a theoretically natural assumption will be the \textit{ground state}. Even if our wave function is not in the ground state, it is still worthwhile to study the case for the ground state, because it can be a good reference for other wave functions. The quantum gravitational version of the ground state wave function was developed by Hartle and Hawking using the Euclidean path integral \cite{Hartle:1983ai}; so-called the no-boundary wave function. The no-boundary wave function can be approximated by sum-over instantons. The instantons can be interpreted as classical histories, and the Euclidean action integral for the instanton gives the probability of each history.

There are a lot of papers regarding the no-boundary wave function. However, almost all of them considered the slow-rolling fields. If the field is no more slowly rolling, then there is no way to deal with the instanton by an analytic method. On the other hand, if we analyze the instantons by a numerical technique \cite{Hartle:2007gi,Hartle:2008ng}, then we can deal with more general and realistic potentials. In general, instantons should be complexified and eventually we will require the classicality for the complexified instanton \cite{Halliwell:1989dy,Halliwell:1990qr,Barvinsky:1993ne}. After finding the classicalized instantons using a numerical searching algorithm (we follow the authors' previous work \cite{Hwang:2012mf,Hwang:2011mp,Hwang:2012zj,Hwang:2012bd}), we can obtain the probability distribution for the initial condition of the universe in a general potential, without assuming slow-roll conditions.

In this paper, we apply the no-boundary wave function to various inflation models. We know that traditionally the result was negative; the no-boundary wave function seems not prefer large $e$-foldings \cite{Hawking:1984hk} and hence we need further assumptions or we need to find an alternative to inflation. However, by extending our advanced methods for dynamical instantons and by applying various ideas of models, we may find a good inflation model that is compatible with the no-boundary wave function. In addition, our study can answer some of traditional doubts on inflation.

This paper is organized as follows: in Section~\ref{sec:pre}, we discuss important ingredients on inflation and Euclidean quantum cosmology. In Section~\ref{sec:nbm}, we first define a quantitative standard to assess model hypothesis (typicality) and illustrate three possible ways toward a good hypothesis. Finally, in Section~\ref{sec:dis}, we summarize and give implications.

\section{\label{sec:pre}Preliminaries}

In this section, we summarize basics which are already known. First, we summarize observables and constraints of inflation that is based on the recent Planck data. Second, to answer on the initial conditions of the inflation, we use the no-boundary wave function; we summarize basics of the no-boundary proposal and complex-valued instantons. Third, we illustrate traditional doubts on inflation. Although some of the doubts can be answered by the no-boundary proposal, the others may not; hence, we illustrate the motivation of this paper to find an inflation model that is suitably compatible with the no-boundary proposal.

\subsection{Observational constraints of inflation models}

We first summarize observational constraints for a single field inflation model. We begin with the Einstein gravity with a single scalar field:
\begin{eqnarray}\label{eq:ac}
S = \int \sqrt{-g} dx^{4} \left[\frac{1}{16\pi}R - \frac{1}{2} \left(\nabla \phi \right)^{2} - V(\phi) \right].
\end{eqnarray}
We use the Planck unit such that $G = c = \hbar = 1$. In addition, to compare with physical dimensions, we introduce the reduced Planck mass: $M_{\mathrm{P}} = 1/\sqrt{8\pi}$. We impose the following metric ansatz to describe a homogeneous and isotropic universe:
\begin{eqnarray}
ds^{2} = - dt^{2} + \rho^{2}\left(t\right) d\Omega_{3}^{2}.
\end{eqnarray}

In a single field inflation model, the equation of motion of inflaton is
\beq
0 = \ddot{\phi} + 3 H \dot{\phi} + V',
\eeq
where the derivatives $\dot{}$ and $'$ are with respect to $t$ and $\phi$ respectively, and $H = \dot{\rho}/\rho$. Furthermore, in the slow-roll approximation,
\beq
0 \approx 3 H \dot{\phi} + V'.
\eeq

The observational parameters are summarized as follows. The slow-roll parameters are defined as \cite{slowroll}
\beq
\epsilon \equiv \frac{1}{2} \left( \frac{M_\p V'}{V} \right)^2, \quad \eta \equiv \frac{M_\p^2 V''}{V}.
\eeq
The power spectra of the field fluctuations are
\beq
\mathcal{P}_\phi = \left( \frac{H}{2 \pi} \right)^2, \quad \mathcal{P}_{+,\times} = \frac{2}{M_\p^2} \left( \frac{H}{2 \pi} \right)^2.
\eeq
The corresponding curvature perturbation is given by $\mathcal{R} = \left( - H / \dot{\phi} \right) \delta \phi$ (valid in linear perturbation theory) \cite{Sasaki:1995aw}.
The curvature and tensor perturbations are given by
\bea
\delta_H^2 \equiv \frac{4}{25} \mathcal{P}_{\mathcal{R}} &\approx& \frac{1}{75 \pi^2 M_\p^6} \frac{V^3}{V'^2} = \frac{1}{150 \pi^2 M_\p^4} \frac{V}{\epsilon},
\\
\mathcal{P}_T = 2 \mathcal{P}_{+,\times} &=& \frac{V}{3 \pi^2 M_\p^4}.
\eea
The spectral indices are defined as
\bea
n_s - 1 &\equiv& \frac{ d \mathcal{P}_{\mathcal{R}}}{d \log \tilde{k}} \approx 2 \eta - 6 \epsilon,
\\
n_T &\equiv& \frac{d \mathcal{P}_T}{d \log \tilde{k}} \approx - 2 \epsilon.
\eea
The tensor-to-scalar ratio is defined as
\beq
r \equiv \frac{\mathcal{P}_T}{\mathcal{P}_{\mathcal{R}}} \approx 16 \epsilon.
\eeq
The number of $e$-folding is given by
\beq
\mathcal{N}_* = \int_{t_*}^{t_{\rm e}} dt H \approx - \frac{1}{M_\p^2} \int_{\phi_*}^{\phi_{\rm e}} d \phi \frac{V}{V_\phi}.
\eeq
The far right-hand side is from the slow-roll approximation.
For a comoving scale $\tilde{k}_*$, we need
\beq
\mathcal{N}_* \approx 71.21 - \log \left( \frac{\tilde{k}_*}{a_0 H_0} \right) + \frac{1}{4} \log \left( \frac{V_{\rm hor}}{M_\p^4} \right) + \frac{1}{4} \log \left( \frac{V_{\rm hor}}{\rho_{\rm end}} \right) + \frac{1 - 3 w}{12 ( 1 + w)} \log \left( \frac{\rho_{\rm R}}{\rho_{\rm end}} \right)
\eeq
where $w$ is the effective equation of state between the end of inflation and reheating.
$\rho_{\rm R}$, the energy density of the universe at reheating, is quite uncertain since it depends on how inflaton couples to visible sector fields.
However, it is lower-bounded to $\left[\mathcal{O}(10) {\rm MeV} \right]^4$ for the successful big bang nucleosynthesis.
Hence there is about $10$ variation of $\mathcal{N}_*$ at most.

The parameterization by Planck Team \cite{Ade:2013uln} is
\bea
\mathcal{P}_{\mathcal{R}} &=& A_s \left( \frac{\tilde{k}}{\tilde{k}_0} \right)^{n_s - 1 + \frac{1}{2} (d n_s / d \log \tilde{k}) \log (\tilde{k}/\tilde{k}_0) + \dots},
\\
\mathcal{P}_T &=& A_T \left( \frac{\tilde{k}}{\tilde{k}_0} \right)^{n_T + \frac{1}{2}(d n_s / d \log \tilde{k}) \log (\tilde{k}/\tilde{k}_0) + \dots},
\eea
where
\bea
A_s &\approx& \frac{1}{24 \pi^2 M_\p^4} \frac{V}{\epsilon},
\\
A_T &\approx& \frac{2 V}{3 \pi^2 M_\p^4},
\eea
and $\tilde{k}_0 = 0.005 {\rm Mpc}^{-1}$ is the pivot scale.

Under the assumption of no spectral running, Planck+WP data implies
\bea
A_s &=& 2.196_{-0.0060}^{+0.0051} \times 10^{-9},
\\
n_s &=& 0.9603 \pm 0.0073.
\eea
At the pivot scale, $\tilde{k}_* = 0.002 {\rm Mpc}^{-1}$, it is constrained as\footnote{In addition, the recent report from BICEP2 collaboration is $0.15 < r_* < 0.27$ \cite{bicep2}.}
\beq
r_* < 0.12.
\eeq
This provides an upper-bound on the energy scale of inflation, $V$,
\beq
V_* \approx \frac{3 \pi^2}{2} A_s r M_\p^4 = \left( 1.94 \times 10^{16} {\rm GeV} \right)^4 \frac{r_*}{0.12}.
\eeq


\subsection{Introduction to the no-boundary proposal}

In this paper, we use the no-boundary wave function \cite{Hartle:1983ai,Hartle:2007gi,Hartle:2008ng} to illustrate the initial conditions of the universe.

\paragraph{The ground state wave function} We begin with the action in Equation~(\ref{eq:ac}). The ground state wave function of the universe \cite{Hartle:1983ai}, or so-called the no-boundary wave function, is given by
\begin{eqnarray}
\Psi[h_{\mu\nu}, \chi] = \int_{\partial g = h, \partial \phi = \chi} \mathcal{D}g\mathcal{D}\phi \;\; e^{-S_{\mathrm{E}}[g,\phi]},
\end{eqnarray}
where $h_{\mu\nu}$ and $\chi$ are the boundary values of the Euclidean metric $g_{\mu\nu}$ and the matter field $\phi$, respectively. The integration is over all non-singular geometries with this given single boundary. After the Wick rotation, we obtain the Euclidean action
\begin{eqnarray}
S_{\mathrm{E}} = -\int d^{4}x \sqrt{+g} \left( \frac{1}{16\pi}R - \frac{1}{2} (\nabla \phi)^{2} - V(\phi) \right).
\end{eqnarray}

\paragraph{Minisuperspace model} In the minisuperspace model by imposing the $O(4)$ symmetry
\begin{eqnarray}
ds_{\mathrm{E}}^{2} = d\tau^{2} + \rho^{2}\left(\tau\right) d\Omega_{3}^{2},
\end{eqnarray}
the Euclidean action is reduced by
\begin{eqnarray}
S_{\mathrm{E}} = 2 \pi^{2} \int d\tau \left[ -\frac{3}{8\pi} \left( \rho \dot{\rho}^{2} + \rho \right) + \frac{1}{2}\rho^{3} \dot{\phi}^{2} + \rho^{3} V(\phi) \right],
\end{eqnarray}
where the dot represents the derivation with respect to the Euclidean time. For numerical convenience, we define the rescaling without changing dynamics:
\begin{eqnarray}\label{eq:rescalings}
d\tau = \frac{d\tilde{\tau}}{\sqrt{V_{0}}}, \;\;\;\; \rho = \frac{\tilde{\rho}}{\sqrt{V_{0}}},
\end{eqnarray}
where $V_{0}$ can be an arbitrary constant.
Then we obtain the action:
\begin{eqnarray}
S_{\mathrm{E}} = \frac{\tilde{S}_{\mathrm{E}}}{V_{0}},
\end{eqnarray}
where
\begin{eqnarray}
\tilde{S}_{\mathrm{E}} = 2 \pi^{2} \int d\tilde{\tau} \left[ -\frac{3}{8\pi} \left( \tilde{\rho} \dot{\tilde{\rho}}^{2} + \tilde{\rho} \right) + \frac{1}{2}\tilde{\rho}^{3} \dot{\phi}^{2} + \tilde{\rho}^{3} \frac{V(\phi)}{V_{0}} \right],
\end{eqnarray}
and now the dot is the derivation with respect to $\tilde{\tau}$. In the followings, we calculate $\tilde{S}_{\mathrm{E}}$ numerically for several potentials, and change the typical energy scales of inflation by varying $V_{0}$.

\paragraph{Steepest-descent approximation} We use the steepest-descent approximation and only consider the on-shell solutions to count probability from the path-integral \cite{Hartle:1983ai,Hartle:2007gi,Hartle:2008ng}. We solve the classical equations of motion for Euclidean and Lorentzian time directions
\begin{eqnarray}
\label{eq:1}\ddot{\phi} &=& - 3 \frac{\dot{\rho}}{\rho} \dot{\phi} \pm V',\\
\label{eq:2}\ddot{\rho} &=& - \frac{8 \pi}{3} \rho \left( \dot{\phi}^{2} \pm V \right),
\end{eqnarray}
where the upper sign is for the Euclidean time and the lower sign is for the Lorentzian time.
Note that, due to the Wick rotation of the time, all the functions $\rho$ and $\phi$ are in principle complex valued \cite{Halliwell:1989dy,Halliwell:1990qr}. By imposing the equations of motion, we obtain the on shell action and this becomes
\begin{eqnarray}
S_{\,\mathrm{E}} = 4\pi^{2} \int d \tau \left( \rho^{3} V - \frac{3}{8 \pi} \rho \right).
\end{eqnarray}
In general we have to impose eight initial conditions: real part and imaginary part of $\rho$, $\dot{\rho}$, $\phi$, $\dot{\phi}$. However, we should fix six of them for a compact and regular manifold:
\begin{eqnarray}\label{eq:initial}
\rho_{\tau=0} = 0, \;\;\; \dot{\rho}_{\tau=0} = 1, \;\;\; \dot{\phi}_{\tau=0} = 0.
\end{eqnarray}
Hence, two conditions for $\phi_{\tau=0}$ are still remained. In addition, we want to analytically continue to the Lorentzian geometry using $\tau = X + i t$. Because of the analyticity, at the turning point $\tau = X$, we impose
\begin{eqnarray} \label{eq:paste}
\rho_{t=0} = \rho_{\tau=X}, \;\; \dot{\rho}_{t=0}=i\dot{\rho}_{\tau=X}, \;\; \phi_{t=0} = \phi_{\tau=X}, \;\; \dot{\phi}_{t=0}=i\dot{\phi}_{\tau=X}.
\end{eqnarray}

\paragraph{Classicality} Because of the analytic continuation to complex functions, the action is in general complex \cite{Hartle:2007gi,Hartle:2008ng,Halliwell:1989dy,Halliwell:1990qr}, so that
\begin{equation}
\Psi[a,\chi] \simeq A[a,\chi] e^{i S[a,\chi]},
\label{eq:class}
\end{equation}
where $a$ and $\chi$ are boundary values of $\rho$ and $\phi$, respectively, and $A$ and $S$ are real. If the rate of change of $S$ is much greater than that of $A$,
\begin{equation} \label{eqn:classicality}
\left|\nabla_I A\left[a, \chi\right]\right|\ll \left|\nabla_I S\left[a, \chi\right]\right|, \qquad I=a, \chi,
\end{equation}
then the wave function describes almost classical behavior.

We require initial conditions as Equation~(\ref{eq:initial}) for regularity and, at the junction time $\tau = X$, we paste $\rho(\tau)$ and $\phi(\tau)$ to $\rho(t)$ and $\phi(t)$ as Equations~(\ref{eq:paste}). The remaining initial conditions are the initial field value $\phi(0) = \phi_{0} e^{i\theta}$, where $\phi_{0} \geq 0$ is a modulus and $\theta$ is a phase angle. For a fixed $\phi_{0}$, we tune two parameters $\theta$ and the turning point $X$ to satisfy the classicality condition \cite{Hartle:2007gi,Hartle:2008ng}. If we can find such an instanton, then we interpret it as a history of the universe created from nothing with the probability
\begin{equation}
P \propto e^{- 2 S_{\mathrm{E}}}.
\end{equation}

\paragraph{General properties and comments} We summarize general properties of the no-boundary wave function in previous literature.

\begin{enumerate}
\item When the mass-potential energy ratio $m^{2}/V(\phi_\mathrm{m})$ of the local minimum $\phi_\mathrm{m}$ is greater than a critical value (greater than $6\pi$), there appears a cutoff on the initial conditions \cite{Hartle:2007gi,Hartle:2008ng}; there is no classicalized history if the initial field value is smaller than the cutoff value. This implies that to become a classicalized history, there should be a time period of experiencing slow-roll inflation.

\item Therefore, for classicalized histories, at the turning point, the scalar field is not so energetic: $\dot{\phi}^{2} \ll V$ \cite{Hartle:2007gi,Hartle:2008ng}.

\item In the slow-roll limit, the probability is approximately $\log P \simeq - 2 S_{\mathrm{E}} \simeq 3/8V(\phi_{\tau=0})$, since the solution is quite similar to the Hawking-Moss instanton \cite{Hawking:1981fz}. In addition, $\phi_{\tau=0} \simeq \phi_{t=0}$ \cite{Hartle:2007gi,Hartle:2008ng,Lyons:1992ua}.

\item In the slow-roll limit with a steeper potential with dynamical scalar field, we still can rely on $\log P \simeq 3/8V(\phi_{t=0})$ as a good description, while $\phi_{\tau=0} \neq \phi_{t=0}$ \cite{Hwang:2012mf}. The probability of each classical history is fixed during the time evolution. However, the field value is changed along the history. Then, the probability should be presented as a function of the field value $\phi$ and time $t$, or equivalently a canonical variable $\rho$. In this limit, $\rho$ corresponds the wavelength cutoff of the quantum field theoretical approach \cite{Hwang:2012mf}.

\item In the non-slow-roll limit, different features arise from the highly dynamical scalar field. For example, around a steep local maximum, there is a tendency that the probability becomes flat \cite{Hwang:2011mp,Hwang:2012zj}.
\end{enumerate}

In this paper, we denote $\phi_{\mathrm{top}}$ as the local maximum of the potential or the boundary of the valid region of the effective potential (and we choose $\phi_{\mathrm{top}}=0$ without loss of generality) and $\phi_{\mathrm{m}}$ as the local minimum. In addition, we assume that the potential is monotone between $\phi_{\mathrm{top}}$ and $\phi_{\mathrm{m}}$. Because of the fourth property, it is convenient to plot the probability as a function of the field at the turning point, $\phi_{t=0}$. Hence, to consider the Lorentzian evolution of Euclidean instantons, the following set of initial conditions is a good choice:
\begin{eqnarray}\label{eq:ini}
\rho(0) = \sqrt{\frac{3}{8\pi V(\phi(0))}}, \;\; \dot{\rho}(0) = 0, \;\; \phi(0)=\phi_{\mathrm{m}} 2^{-k}, \;\; \dot{\phi}(0)=0,
\end{eqnarray}
where $\rho(0)$ and $\dot{\rho}(0)$ are familiar forms of the Hawking-Moss instanton and $\dot{\phi}(0)$ follows from the second property. $\phi(0)$ is chosen to approach to the origin, $\phi_{\mathrm{top}}$, as $k$ increases. To quantify the number of $e$-foldings during inflation, we estimate
\begin{eqnarray}
\mathcal{N} = \log \frac{\rho(t_{\mathrm{end}})}{\rho(0)},
\end{eqnarray}
where $t_{\mathrm{end}}$ is the time when inflation ends as $\dot{\phi}^{2} \sim V(\phi)$. If we choose the time $t_{\mathrm{end}}$ when the field touches the local minimum $\phi_{\mathrm{m}}$, then it will be a good signature of the beginning of the matter dominated era, since the universe just begins to oscillate around the local minimum.

The first and the third property imply that although the no-boundary wave function requires the existence of inflationary era, it does not prefer large $e$-foldings in general. Rather, it exponentially prefers small $e$-foldings. Hartle, Hawking and Hertog \cite{Hartle:2007gi,Hartle:2008ng} proposed an idea to resolve this. They weighted the volume of the past history to the wave function:
\begin{equation}\label{eq:HHH}
P_{\mathrm{HHH}} \propto e^{- 2 S_{\mathrm{E}} + 3 \mathcal{N}}.
\end{equation}
In this paper, we suggest other possible resolutions (for a previous work, see \cite{Hwang:2012bd}) and criticize the volume weighting.

\subsection{\label{sec:dob}Doubts on inflation}

To find a good inflationary model, we require the following three conditions:
\begin{itemize}
\item[1.] Is the model consistent with the \textit{observational constraints}?
\end{itemize}
Needless to say, this is a basic requirement.
\begin{itemize}
\item[2.] To explain our universe with observational constraints, does the required set of initial conditions \textit{typical} in terms of the no-boundary wave function?
\end{itemize}
If we trust that the no-boundary proposal gives the proper initial condition of our universe as a \textit{working hypothesis}, then the model should explain our universe with a reasonable probability.
In this paper, we give a \textit{quantitative standard} to measure the typicality (Section~\ref{sec:typ}).
\begin{itemize}
\item[3.] Is the model \textit{natural}, i.e., does the model have less fine-tunings and unnatural assumptions? And/or, can it be derived from string theory or more fundamental theory of quantum gravity?
\end{itemize}
A fine-tuned model with exotic assumptions would easily satisfy the first two conditions. However, it may not be regarded as a good model.

On the other hand, the previous conditions are related to the traditional doubts on inflation.
\begin{itemize}
\item[2$'$.] On initial conditions and typicality:
\begin{itemize}
\item[A.] \textit{The no-boundary wave function does not prefer large $e$-folding of inflation} \cite{Hawking:1984hk,Dyson:2002pf}: Hence, large $e$-foldings are thermodynamically or entropically disfavored. Perhaps, we need to seek more direct explanation from the no-boundary proposal, or assume some additional weighting \cite{Hawking:2002af,Hartle:2007gi,Hartle:2008ng}, or assume that our universe was not in the ground state \cite{Vilenkin:1986cy}, or consider an alternative to inflation \cite{Khoury:2001wf}.
\item[B.] \textit{The multiverse measure problem}, if the initial conditions came from the multiverse \cite{Aguirre:2006ak}: Although large $e$-foldings are not preferred by the wave function, one may easily find an inflationary region because it has large volume. However, the volume-weighting is not well-defined and causes difficult problems \cite{Dyson:2002pf}. To explain our universe, we may not rely on multiverse, or we need to directly explain the consistent measure over the multiverse \cite{Linde:2006nw}.
\item[C.] \textit{Convexity of the potential} \cite{Ijjas:2013vea}: There is no sufficient reason why a convex part of potential is preferred over a concave part. This doubt is caused after the Planck data is released.
\end{itemize}
\item[3$'$.] On the naturalness of the model/potential:
\begin{itemize}
\item[D.] For some inflation models, the super-Planckian field values are required. Of course, if the model can be justified from more fundamental theory, e.g., string theory, then we are free from this problem.
\item[E.] For all inflation models, the potential should be finely tuned to satisfy the slow-roll conditions.
\end{itemize}
\end{itemize}

The no-boundary proposal resolves doubt B since it follows the many-world interpretation.
In this paper, we introduce a \textit{quantitative standard} which tests the likelihood of large inflation (Section~\ref{sec:typ}).
It helps finding possible ideas that explain A and C (Sections~\ref{sec:ta}, \ref{sec:bo}, \ref{sec:ch}).
Then, we will check the naturalness in terms of the doubts D and E, by comparing strong points and weak points of each idea (Section~\ref{sec:dis1}).
Ultimately, it will give us a general feeling whether inflation with the no-boundary proposal is a good hypothesis to explain our observed universe.

\section{\label{sec:nbm}No-boundary wave function for inflation models}

In this section, we introduce various inflation models and apply the no-boundary wave function.
We first comment on the curvaton field \cite{Lyth:2001nq} to explain the reason why we focus on the $e$-foldings over other observables.
Let us consider that the matter Lagrangian is as follows:
\beq
\mathcal{L}_{\mathrm{M}} = \frac{1}{2} \left( \nabla \phi \right)^2 + \frac{1}{2} \left( \nabla \psi \right)^2 - V(\phi) - U(\psi)
\eeq
with $\phi$ and $\psi$ being the inflaton and curvaton field, respectively.
Here, we give the curvaton potential $U(\psi)$ assumed to be
\begin{eqnarray}
U(\psi) = \Lambda^4 \left[ 1 - \cos \left( \frac{\psi}{f_\psi} \right) \right] = \Lambda^4 + \frac{1}{2} \left( \frac{\Lambda^2}{f_\psi} \right)^2 \psi^2 + \dots
\end{eqnarray}
As long as the curvaton mass is sufficiently smaller than the mass scale of inflaton, we can ignore the dynamics of the curvaton when we assign probability from the no-boundary proposal because the curvaton field slowly rolls until the instanton is classicalized and the probability is determined. Hence, effectively, the curvaton only shifts cosmological constant so that
\begin{equation}
\log P(\phi_{t=0},\psi_{t=0}) \simeq \frac{1}{V_{0} + U(\psi_{t=0})},
\end{equation}
where we regard $V_{0} \gg U(\psi)$ with the typical energy scale $V_{0}$ of the inflaton $\phi$. The no-boundary wave function does not sensitively depend on the initial position or the dynamics of curvaton field.
Therefore, for a given inflation model, we can give a burden of density perturbations to the curvaton field without changing the results from the no-boundary wave function significantly.
For that reason, we mainly focus on \textit{$e$-foldings} as a candidate to apply the no-boundary proposal.

\subsection{\label{sec:typ}Typicality: the condition for good hypothesis}

Usually, the probability will be calculated by
\begin{eqnarray}
P[\mathcal{N} \geq \mathcal{N}_{*}] &=& \int_{\mathcal{N}[\phi]\geq\mathcal{N}_{*}} P[\phi] d\mu[\phi]\\
&\simeq& \frac{\int_{\mathcal{N}[\phi]\geq\mathcal{N}_{*}} e^{-2S_{\mathrm{E}}[\phi]} d\mu[\phi]}{\int_{\mathcal{N}[\phi]\geq\mathcal{N}_{*}} e^{-2S_{\mathrm{E}}[\phi]} d\mu[\phi] + \int_{\mathcal{N}[\phi] < \mathcal{N}_{*}} e^{-2S_{\mathrm{E}}[\phi]} d\mu[\phi]},
\end{eqnarray}
where we approximate that $P[\phi] \propto \exp -2S_{\mathrm{E}}$ and exponential factors are dominant, as well as we assume that it is reasonable to choose that the field value measure as $d\mu[\phi] = d\phi$.

Since $P[\mathcal{N} \geq \mathcal{N}_{*}] + P[\mathcal{N} < \mathcal{N}_{*}] = 1$, this allows us to define the typicality $\mathcal{T}$ as a standard to determine the good hypothesis:
\begin{eqnarray}
\mathcal{T} \simeq \frac{\int_{\phi_{\mathrm{top}}}^{\phi_{\mathcal{N}_{*}}} \exp \left[ \frac{- 2 \tilde{S}_{\mathrm{E}}}{V_{0}} \right] d\phi}{\int^{\phi_{\mathrm{c}}}_{\phi_{\mathcal{N}_{*}}} \exp \left[ \frac{-2 \tilde{S}_{\mathrm{E}}}{V_{0}} \right] d\phi},
\end{eqnarray}
where $\phi_{\mathrm{c}}$ is the cutoff due to the classicality condition (Equation~(\ref{eqn:classicality})), $\phi_{\mathcal{N}_{*}}$ is the point resulting more than $\mathcal{N}_{*}$ $e$-foldings, and $\phi_{\mathrm{top}}$ is the hill-top. Here, we explicitly separate $\tilde{S}_{\mathrm{E}}$ and $V_{0}$ (the typical energy scale of inflation) to distinguish the role of the energy scale.

Note that
\begin{eqnarray}
\exp \left[-\frac{2\tilde{S}_{\mathrm{E}}[\phi_{\mathrm{top}}]}{V_{0}} \right] \leq P[\phi] \leq \exp \left[ -\frac{2\tilde{S}_{\mathrm{E}}[\phi_{\mathrm{M}}]}{V_{0}} \right],
\end{eqnarray}
where $\phi_{\mathrm{M}}$ ($\simeq \phi_{\mathrm{c}}$, see the next subsection) is the point with maximum Euclidean probability and $\phi_{\mathrm{top}}$ corresponds to the local maximum of the potential (hence, this has the lowest probability). Hence the following is a safe approximation for the typicality estimation:
\begin{eqnarray}
\log \mathcal{T} \gtrsim  - \frac{2}{V_{0}} \left|\tilde{S}_{\mathrm{E}}[\phi_{\mathrm{M}}] - \tilde{S}_{\mathrm{E}}[\phi_{\mathrm{top}}]\right| + \log \frac{\phi_{\mathcal{N}_{*}} - \phi_{\mathrm{top}}}{\phi_{\mathrm{M}}-\phi_{\mathcal{N}_{*}}}.
\end{eqnarray}
In other words, using the parameterization of Equation~(\ref{eq:ini}),
\begin{eqnarray}
\log \mathcal{T} \gtrsim - \frac{2}{V_{0}} \left|\tilde{S}_{\mathrm{E}}[k_{\mathrm{M}}] - \tilde{S}_{\mathrm{E}}[k_{\mathrm{top}}]\right|  + \log \frac{2^{-k_{\mathcal{N}_{*}}}-2^{-k_{\mathrm{top}}}}{2^{-k_{\mathrm{M}}} - 2^{-k_{\mathcal{N}_{*}}}},
\end{eqnarray}
where $k_{\mathrm{M}}$ ($\simeq k_{\mathrm{c}}$, see the next subsection) is the point with maximum Euclidean probability and $k_{\mathrm{top}}$ corresponds to the local maximum of the potential (hence, this has the lowest probability).

This presentation of the typicality is useful, since we obtained a formula that is independent from the normalization factor. If the typicality is close to one, then the hypothesis explains the $\mathcal{N}_{*}$ $e$-foldings; if the typicality is extremely smaller than one, then the hypothesis cannot explain the sufficient $e$-foldings.

\begin{figure}
\begin{center}
\includegraphics[scale=0.5]{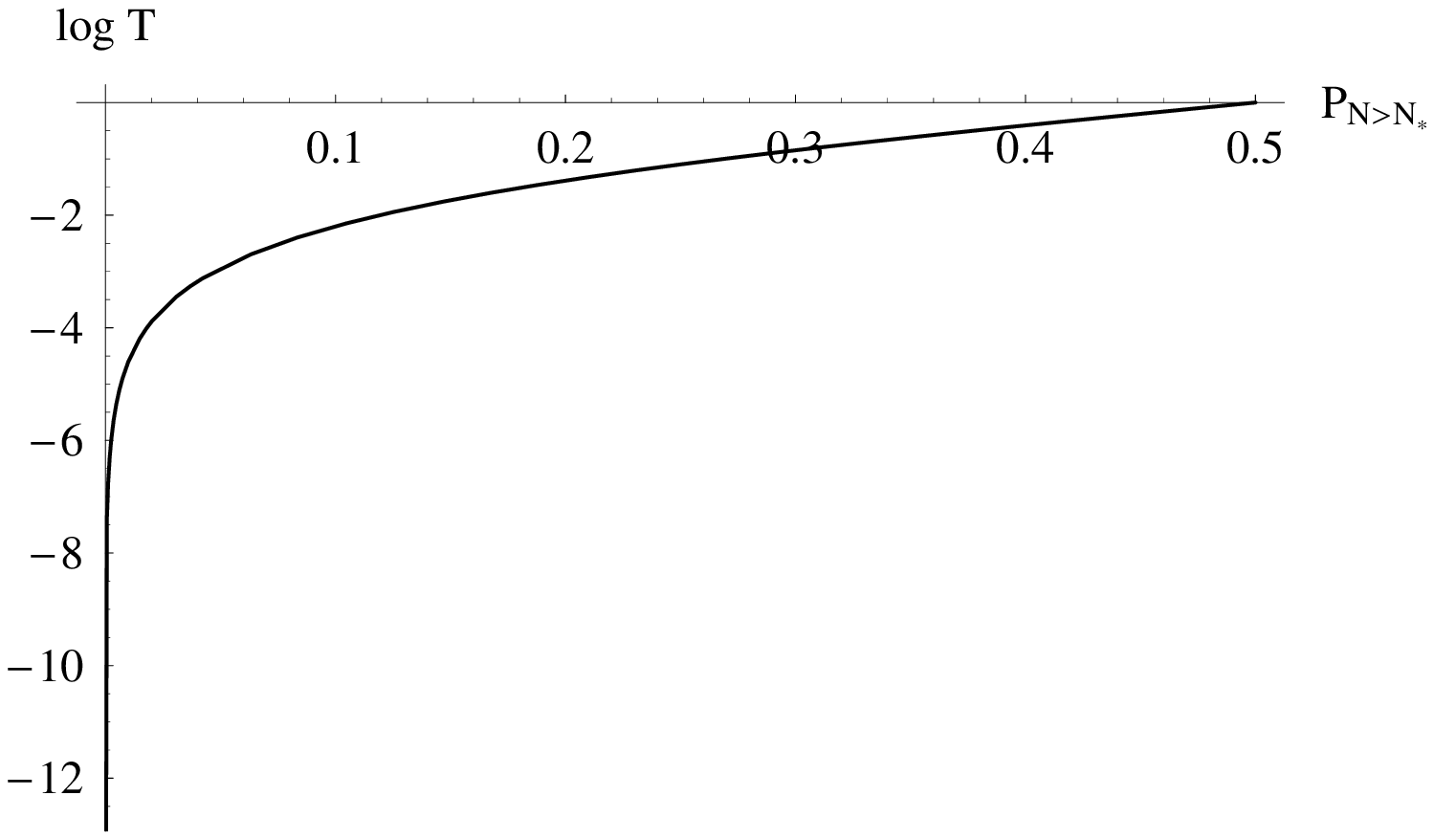}
\caption{\label{fig:LnT}Relation between $P_{\mathrm{cut}}$ and $\log \mathcal{T}_{\mathrm{cut}}$.}
\end{center}
\end{figure}

To be more specific, we need to make a guideline that this typicality can be used to rule out a hypothesis in terms of the probability bound or cutoff (although still there may be no consensus to restrict this probability cutoff), as usual statistical analysis do so. Let us consider the situation that there is a probability cutoff $P_{\mathrm{cut}}$ so that if the probability of more than $\mathcal{N}_{*}$ $e$-foldings is less than $P_{\mathrm{cut}}$, then we rule out the hypothesis. Then, for a given $P_{\mathrm{cut}}$, there is the corresponding $\mathcal{T}_{\mathrm{cut}}$, since $P_{\mathrm{cut}} = \mathcal{T}_{\mathrm{cut}}/(1+\mathcal{T}_{\mathrm{cut}})$. Figure~\ref{fig:LnT} shows the behavior of $\log \mathcal{T}_{\mathrm{cut}}$.
For a given $P_{\mathrm{cut}}$, if the typicality $\log \mathcal{T}$ is greater than $\log \mathcal{T}_{\mathrm{cut}}$, then we say that it is a good hypothesis.

For convenience, we define two quantities:
the action difference
\begin{eqnarray}
\Delta \equiv \left|\tilde{S}_{\mathrm{E}}[\phi_{\mathrm{M}}] - \tilde{S}_{\mathrm{E}}[\phi_{\mathrm{top}}]\right|
\end{eqnarray}
and the field measure difference
\begin{eqnarray}
\Xi \equiv \log \frac{\phi_{\mathcal{N}_{*}}- \phi_{\mathrm{top}}}{\phi_{\mathrm{M}}-\phi_{\mathcal{N}_{*}}}.
\end{eqnarray}
Then the typicality criterion for a good hypothesis becomes
\begin{eqnarray}
-\frac{2}{V_{0}}\Delta + \Xi \gtrsim \log \mathcal{T}_{\mathrm{cut}}.
\end{eqnarray}

Note that the typical energy scale of inflation $V_{0}$ is constrained by observations. For single field inflation, $V_{0} \sim 10^{-7} M_{\mathrm{P}}^{4} \ll M_{\mathrm{P}}^{4}$ is the observational constraint. On the other hand, $\Delta$ and $\Xi$ do not depend on $V_{0}$ in many cases; rather, they only depend on the shape of the potential. Therefore, to build a good model which has our universe as a typical history, there are three possible ways:
\begin{itemize}
\item[P1.] \textit{Pre-inflation near the Planck scale}: Assuming that our universe began with $V_{0} \lesssim \mathcal{O}(1)$ and it could contribute several $e$-foldings. It requires next scenario to explain the observational constraints (Section~\ref{sec:ta}).
\item[P2.] \textit{Tuning potentials}: By tuning the potential, one may reduce $\Delta$ or increase $\Xi$. Some examples are following: small field inflation and scalar-tensor inflation models (Section~\ref{sec:bo}).
\item[P3.] \textit{New ingredient}: If the Euclidean probability has a new ingredient, e.g., $\log P \simeq -2\tilde{S}_{\mathrm{E}}/V_{0} + (\mathrm{contribution\;from\;new\;effects})$, then it may increase the typicality. We discuss this possibility in the context of chaotic inflation and multi-field inflation (Section~\ref{sec:ch}).
\end{itemize}

\subsection{Comments on probability around multiple local extrema}

In the following subsections, we show some examples of probability distribution from the no-boundary proposal. In the paper of Hartle, Hawking, and Hertog \cite{Hartle:2007gi,Hartle:2008ng}, the probability distribution was monotone; largest at the cutoff and monotonically decreases as the potential energy increases. However, if there are multiple local extrema, this is not true in general. For example, if we consider a local maximum and if the mass scale is sufficiently large, then the probability distribution becomes flat \cite{Hwang:2011mp,Hwang:2012zj}.

Furthermore, in this paper, we consider an instanton that moves between the local minimum and the local maximum. Then, all things become complicated. If the field distance between the local minimum and the local maximum is sufficiently large and the potential is sufficiently flat, then we recover the results of the slow-roll approximation \cite{Hartle:2007gi,Hartle:2008ng}. On the other hand, if two extrema are quite close to each other and the curvature around the local maximum is large, then the probability distribution will be flattened again. In this limit, one missing point is the behavior near the cutoff around the local minimum. Will the cutoff be still there, or some new things happen? The answer is that \textit{there is a cutoff, but it may not have the largest probability.}

\begin{figure}
\begin{center}
\includegraphics[scale=0.4]{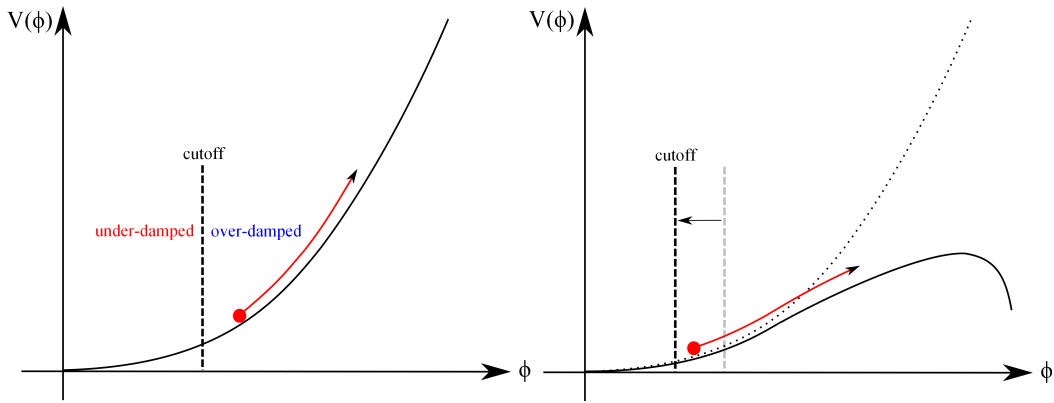}
\caption{\label{fig:prob}If the potential is changed, then the cutoff is shifted, although the local shape of the potential is similar around the cutoff.}
\end{center}
\end{figure}

Figure~\ref{fig:prob} intuitively explains this behavior. Here, red arrows mean the dynamics during the Euclidean time. In left of Figure~\ref{fig:prob}, it shows the quadratic potential case that was studied in \cite{Hartle:2007gi,Hartle:2008ng}. We can easily divide the under-damped region and the over-damped region; the cutoff locates between the two regions. Right of Figure~\ref{fig:prob} shows a potential that is locally (up to the cutoff place of the left figure; gray-dashed line) the same as the quadratic potential. However, the instanton that begins from the left of the gray-dashed line moves toward a region that is quite different from the original quadratic potential. The instanton experiences over-damped inflation in that region. Therefore, even though it begins inside of the previous cutoff, outside region has an effect on its dynamics. This is why the `real' cutoff of the new potential is shifted (black-dashed line).

As we will see in the following subsections, although the cutoff is shifted, the maximum probability is not on the cutoff. Again, this can be understood qualitatively; the region between the gray-dashed line and the black-dashed line depends only on the global shape of the potential and hence it does not necessarily increase the probability as it was in the quadratic potential case. Of course, all the calculations should be confirmed by detailed numerical simulations.

\subsection{Three scenarios toward compatible models}

Regarding the typicality function $\mathcal{T}$, we find inflation models those are compatible with the no-boundary proposal. We propose three possibilities. (P1) The first inflation began with high energy scale: Section~\ref{sec:ta}. (P2) The potential is sufficiently flat to reduce $\Delta$ or the field space of large $e$-foldings is sufficiently wide to increase $\Xi$: Section~\ref{sec:bo}. (P3) The Euclidean probability has a new correction terms: Section~\ref{sec:ch}.

\subsubsection{\label{sec:ta}Inflation around the natural tachyonic top}

In this subsection, we consider the possibility (P1). To investigate this scenario, we first study the characteristics of the Euclidean probability in a tachyonic potential, by varying the mass of the local maximum. Second, we calculate the $e$-foldings and weight the Euclidean probability. This reveals general feeling about the no-boundary proposal with the hill-top potential. Finally, we will discuss a possible scenario that explains sufficient $e$-foldings.

\paragraph{Euclidean probability}

\begin{figure}
\begin{center}
\includegraphics[scale=0.25]{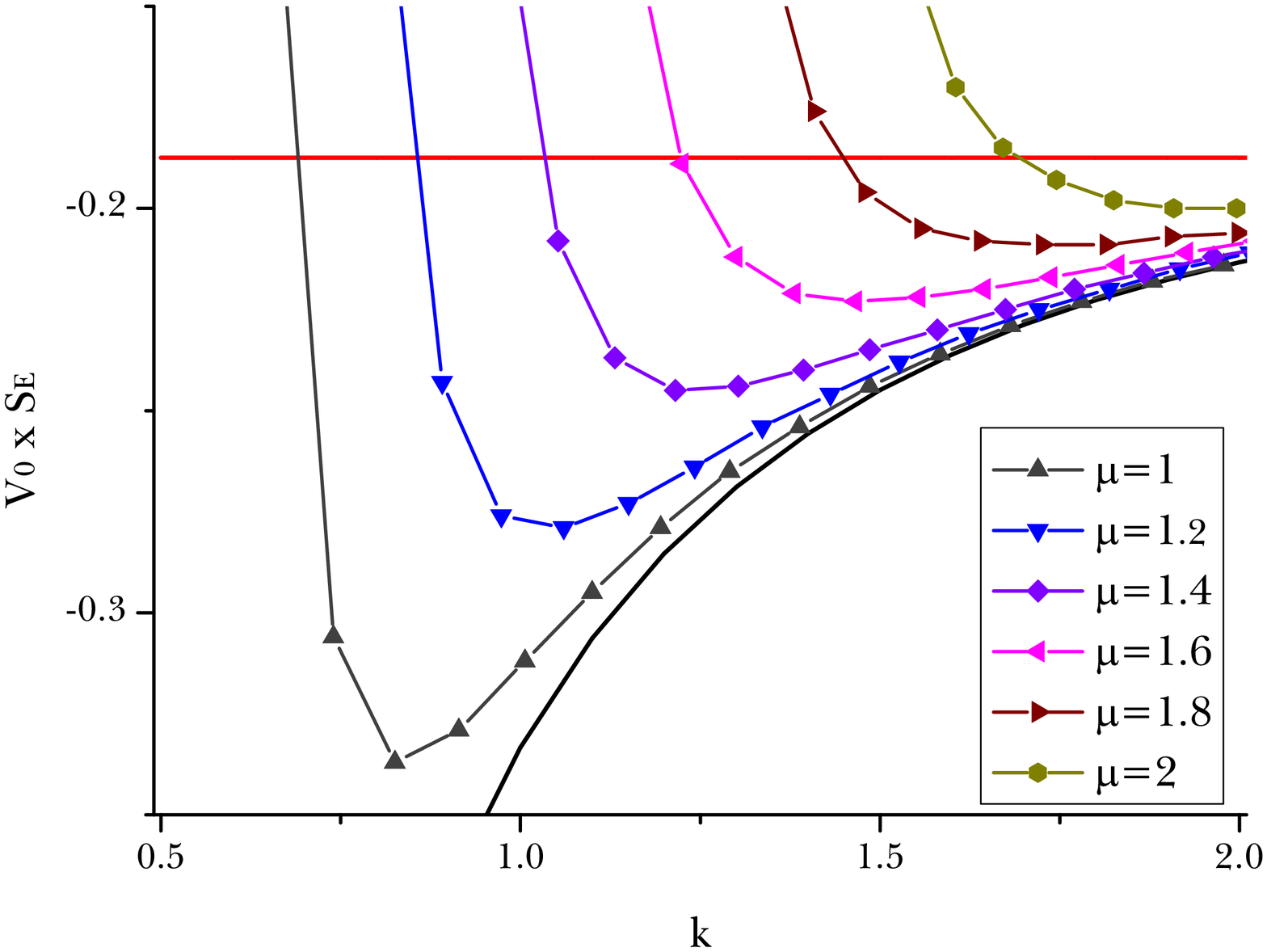}
\includegraphics[scale=0.25]{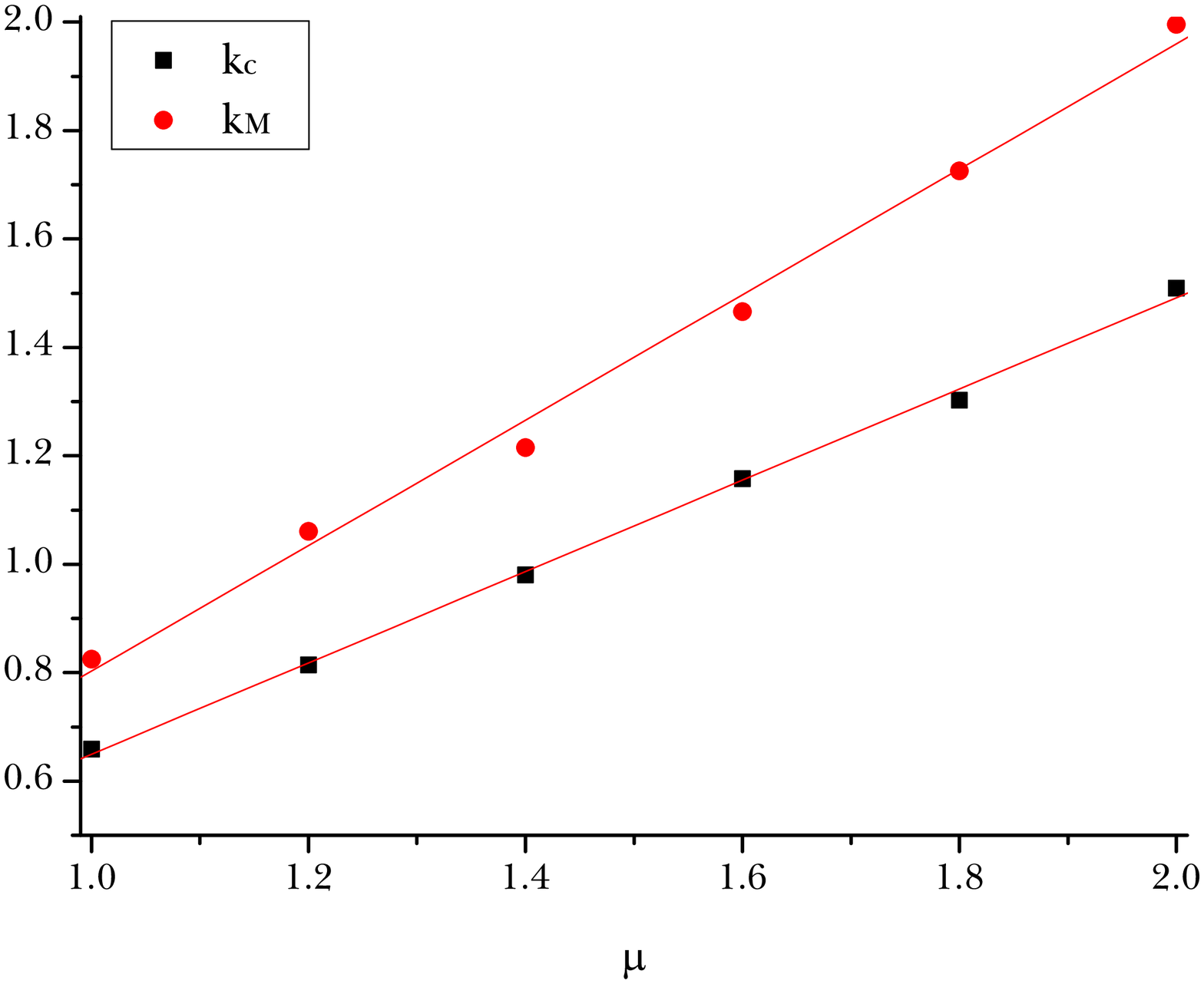}
\caption{\label{fig:probability}Left: The Euclidean action $\tilde{S}_{\mathrm{E}}$ as a function of $k$ by varying $\mu$. The black curve is the analytic fitting curve $-3/16V(\phi)$ and the red curve is the action at the local maximum ($-3/16$). Right: Location of $k_{\mathrm{c}}$ and $k_{\mathrm{M}}$ by varying $\mu$.}
\includegraphics[scale=0.7]{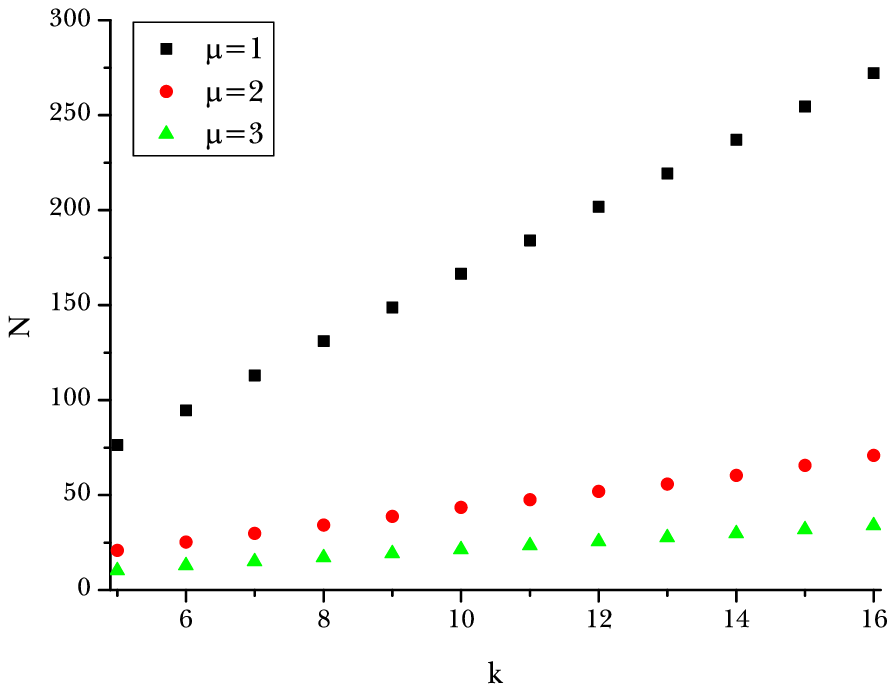}
\includegraphics[scale=0.7]{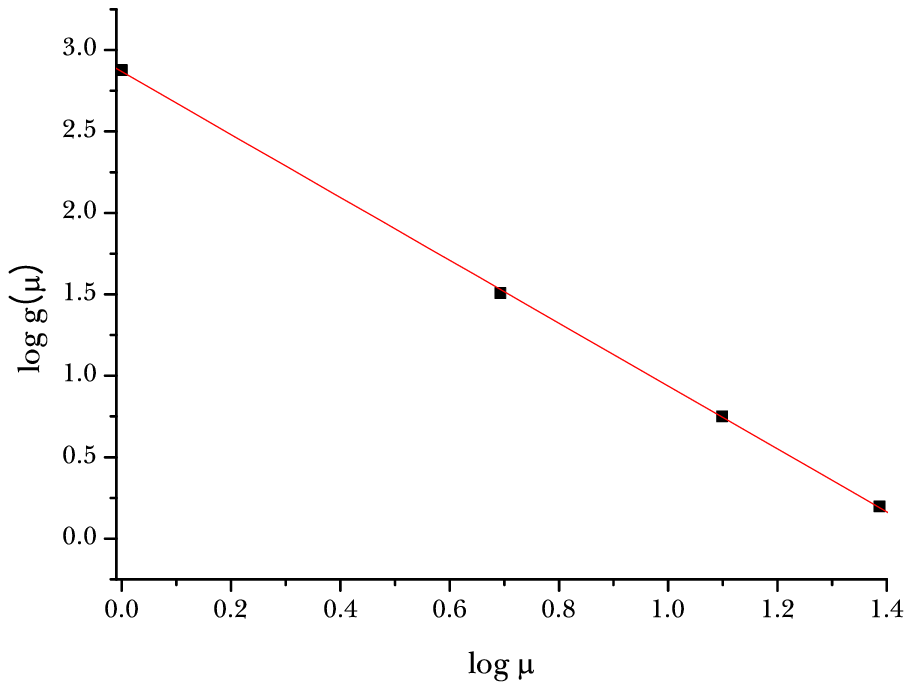}
\caption{\label{fig:efolding_mu}Left: $e$-folding is linearly depend on $k$ (for $\mu=1,2,3$). Right: the function $\log g(\mu)$ is plotted as a function of $\log \mu$ (for $1 \leq \mu \leq 4$). The gradient is $-1.93034$ and the intercept is $2.86726$. Therefore, the constant $C$ is approximately $17.58876$.}
\includegraphics[scale=0.5]{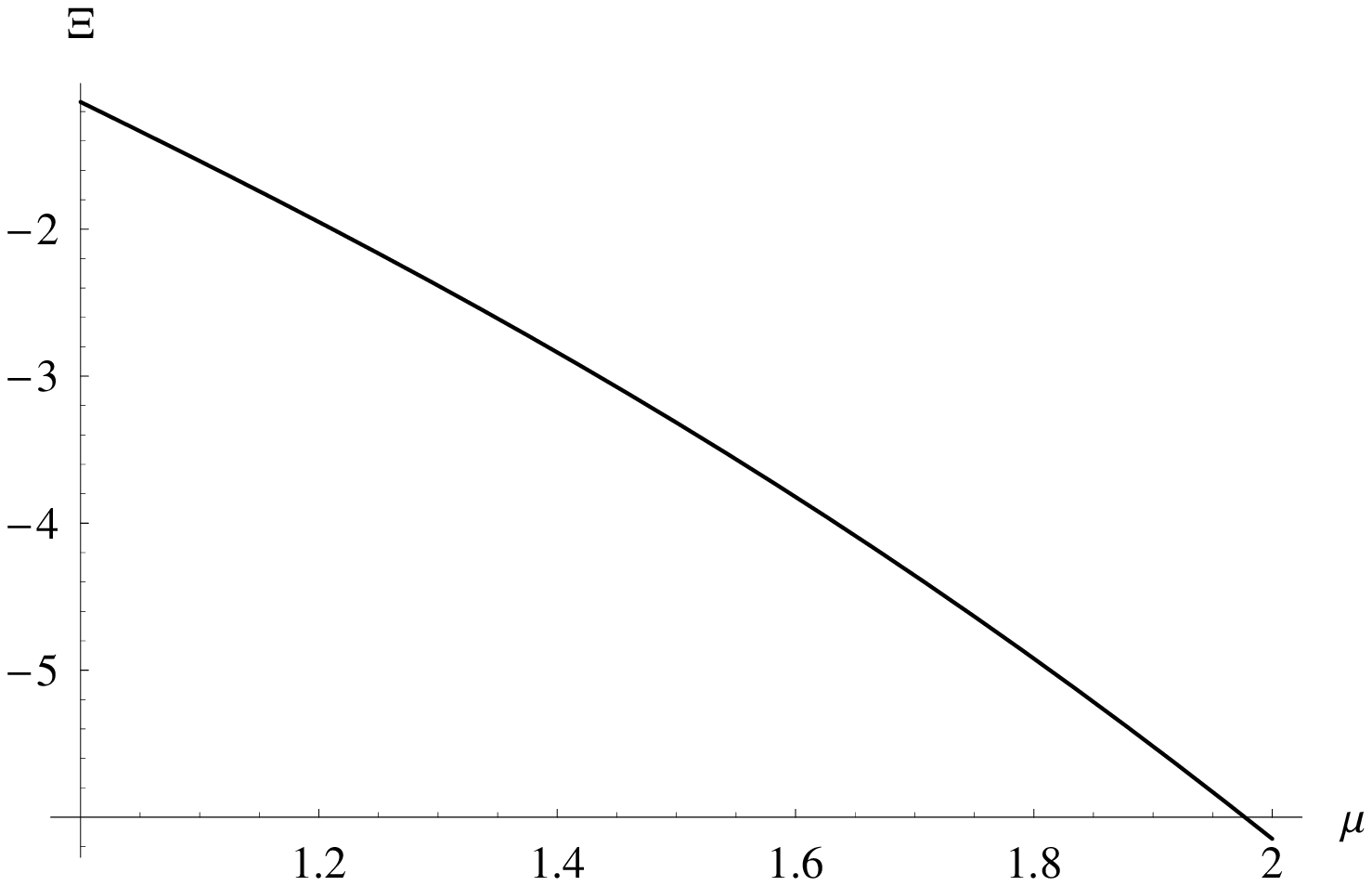}
\caption{\label{fig:Xi_mu} $\Xi$ by varing $\mu$.}
\end{center}
\end{figure}

We use the following tachyonic potential:
\begin{eqnarray}
V(\phi) = V_{0} \left( 1 - \frac{\mu^{2}}{2} \phi^{2} + \frac{\lambda}{4!}\phi^{4} \right),
\end{eqnarray}
where we can choose $V_{0}=1$ due to the re-scaling, Equation~(\ref{eq:rescalings}). Here, we set $\lambda=3\mu^{4}/2$ to be sure $V(\phi_{\mathrm{m}})=0$ where $\pm \phi_{\mathrm{m}}$ is the local minimum of the potential, and $\phi_{\mathrm{m}}=\sqrt{6 \mu^{2}/\lambda}=2/\mu$. In this subsection, we consider $\mu \gtrsim 1$ case and we consider it is natural in some sense. (We calculate for $\mu < 1$ cases, in the next subsection.) By scanning parameter spaces, we obtain Figure~\ref{fig:probability}. This contains information on the Euclidean action and the location of $k_{\mathrm{c}}$ and $k_{\mathrm{M}}$:
\begin{eqnarray}
k_{\mathrm{c}} &\simeq& \beta_{\mathrm{c}} \times \mu + \gamma_{\mathrm{c}},\\
k_{\mathrm{M}} &\simeq& \beta_{\mathrm{M}} \times \mu + \gamma_{\mathrm{M}},
\end{eqnarray}
where for $k_{\mathrm{c}}$, $\beta_{\mathrm{c}} = 0.842$ and $\gamma_{\mathrm{c}} = -0.193$; for $k_{\mathrm{M}}$, $\beta_{\mathrm{M}} = 1.157$ and $\gamma_{\mathrm{M}} = -0.354$.

\paragraph{Number of $e$-foldings}

By varying $\mu$ and $k$, we estimate the number of $e$-foldings as a function of $\mu$ and $k$. Then the following is a good estimation:
\begin{eqnarray}
\mathcal{N} \simeq g(\mu) k,
\end{eqnarray}
where
\begin{eqnarray}
g(\mu) \simeq C \mu^{-p}.
\end{eqnarray}
First, for a given $\mu$, we can show that the number of $e$-foldings linearly increase as a function of $k$ (for example, see Left of Figure~\ref{fig:efolding_mu}). We can estimate the gradient of the linear dependence $g(\mu)$ by varying $\mu$ (Right of Figure~\ref{fig:efolding_mu}). Therefore, we conclude that $C\simeq17.59$ and $p\simeq 1.93$ is a good estimation for the number of $e$-foldings for the fast-roll inflation. The real $e$-foldings are the similar order or slightly larger than this estimation.

\paragraph{Typicality relation and expectation values}

We can calculate
\begin{eqnarray}
k_{\mathcal{N}_{*}} &\simeq& 2.843 \times \mu^{1.93},\\
k_{\mathrm{M}} &\simeq& 1.157 \times \mu - 0.354.
\end{eqnarray}
Hence, $\Xi \sim -\mathcal{O}(1)$ (Figure~\ref{fig:Xi_mu}). In addition, $\Delta \simeq -0.1$ from Figure~\ref{fig:probability}. Therefore, approximately, for $\mu \simeq 1$ and $V_{0}\gtrsim 10^{-2}$, the typicality does not rule out $\mathcal{N}_{*}$ $e$-foldings up to the probability $\sim 1 \%$. On the other hand, if we do not require $\mathcal{N}_{*}$ $e$-foldings for the first inflation, then we can relax the typicality bound.

For further analysis, let us think about the expectation values. Expectation values are not quite meaningful if the probability distribution is not Gaussian, e.g., Equation~(\ref{eq:HHH}); in this case, the standard deviation is the same order of the mean value. On the other hand, if a distribution is similar to Gaussian or the standard deviation is reasonably smaller than the mean value, then the mean value and standard deviation can be meaningful. Let us investigate further on this direction.

\begin{figure}
\begin{center}
\includegraphics[scale=0.4]{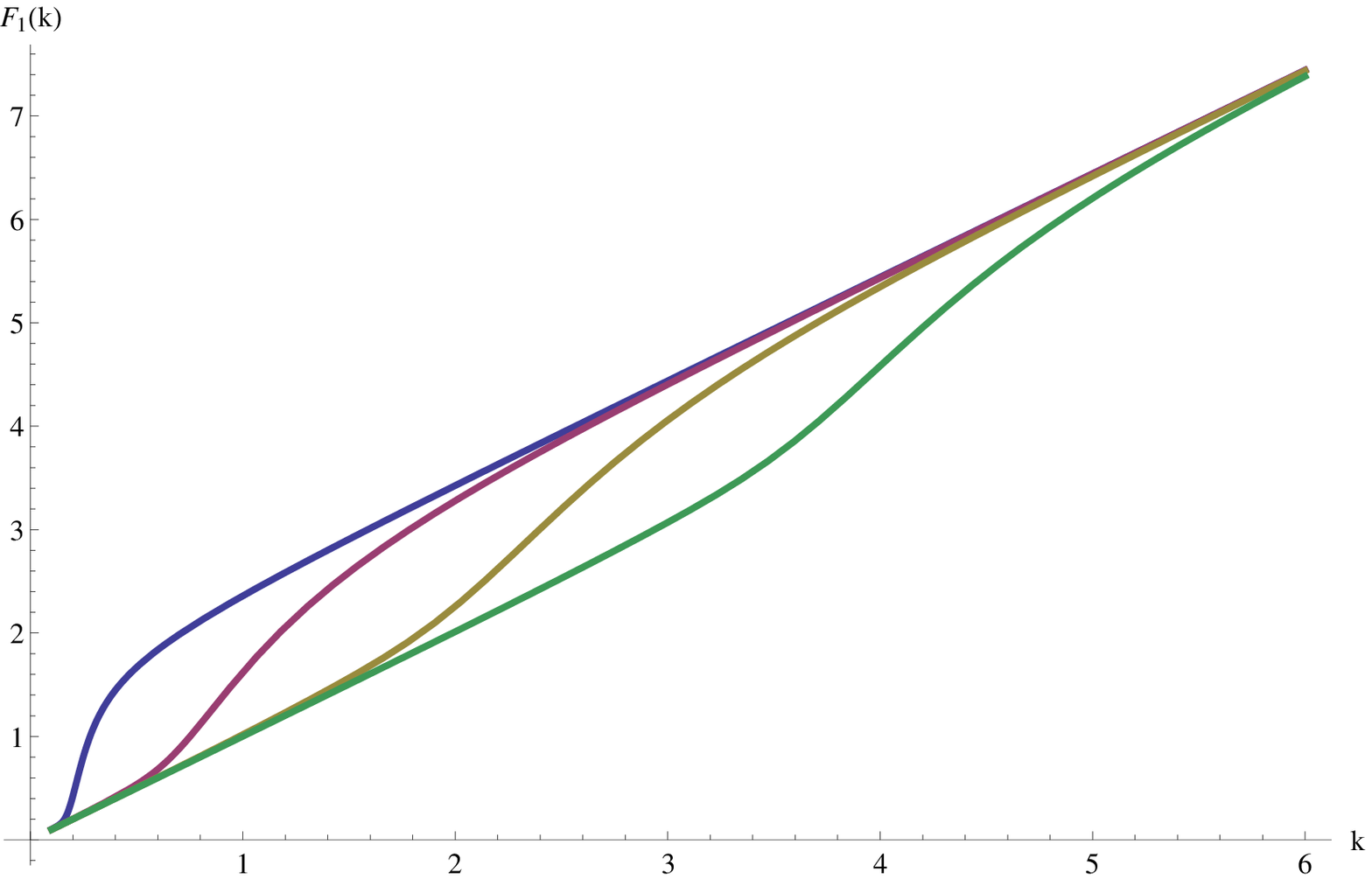}
\includegraphics[scale=0.4]{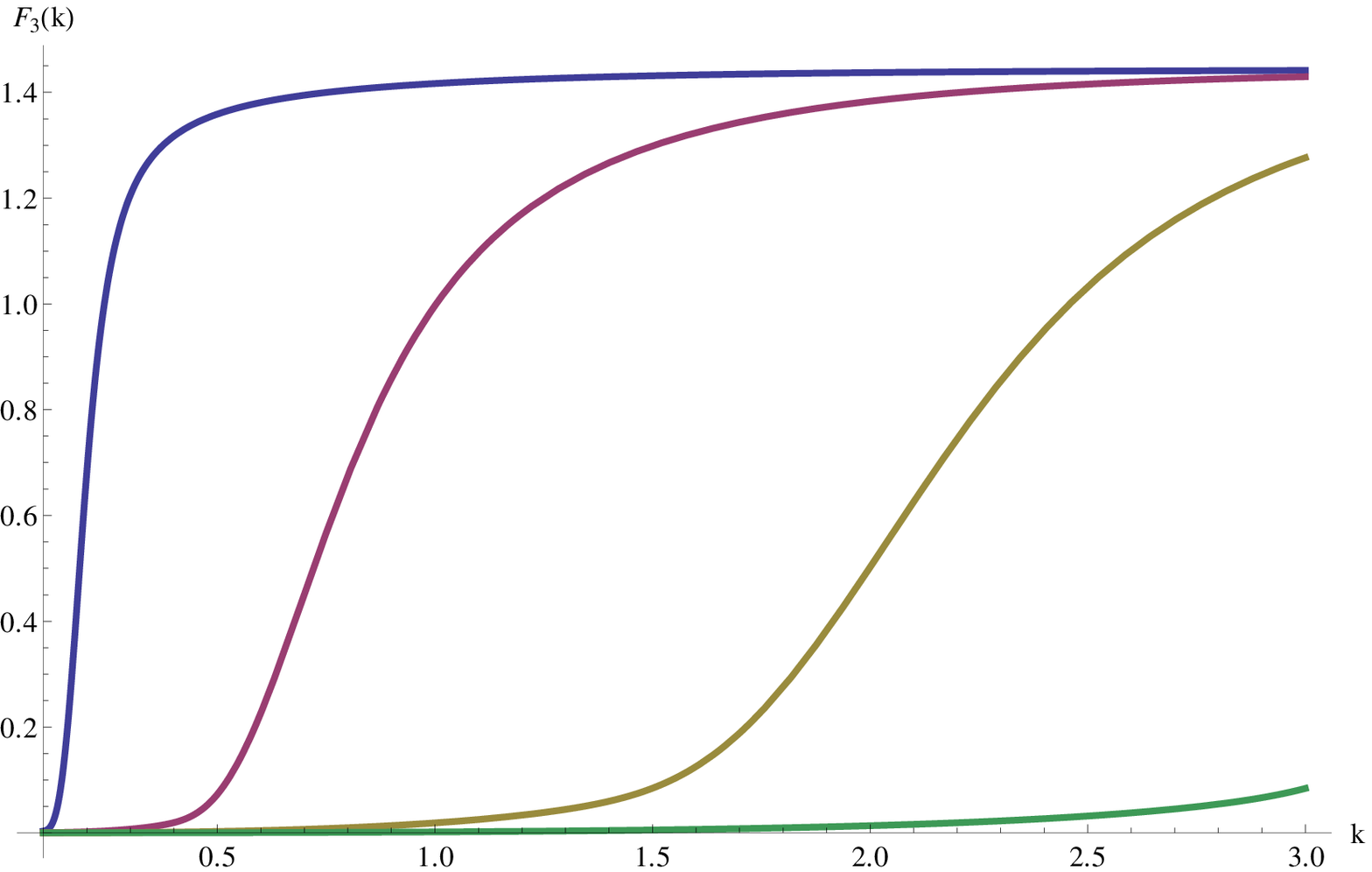}
\caption{\label{fig:F1}$F_{1}(k)$ and $F_{3}(k)$. From left to right (blue, red, yellow, green), $V_{0}=1$, $0.1$, $0.01$, and $0.001$. $F_{1}(k)$ converges to a linear function (the gradient is approximately $1$) and $F_{3}(k)$ converges to a constant (approximately $1.43$), when $k$ is sufficiently large.}
\end{center}
\end{figure}

We introduce useful relations (Left of Figure~\ref{fig:F1}):
\begin{eqnarray}
F_{1} (k,V_{0}) &\equiv& \frac{\int_{0}^{\phi(k)} P(\phi_{0}) k' d\phi_{0}}{\int_{0}^{\phi(k)} P(\phi_{0}) d\phi_{0}}\\
&=& \frac{\int_{k}^{\infty} \exp{\frac{3}{8V_{0}\left( 1 - 2^{-2k'} \right)^{2}}} k' 2^{-k'} dk'}{\int_{k}^{\infty} \exp{\frac{3}{8V_{0}\left( 1 - 2^{-2k'} \right)^{2}}} 2^{-k'} dk'}.
\end{eqnarray}
For special limiting cases, we can estimate more exactly. If $V_{0} \sim 1$, then
\begin{eqnarray}
F_{1} (k,V_{0}) \simeq k + 1.44
\end{eqnarray}
for $k > 0.5$. On the other hand, if $V_{0} \ll 1$, then
\begin{eqnarray}
F_{1} (k,V_{0}) \simeq k.
\end{eqnarray}
In addition, we define that
\begin{eqnarray}
F_{2} (k,V_{0}) &\equiv& \frac{\int_{0}^{\phi(k)} P(\phi_{0}) k'^{2} d\phi_{0}}{\int_{0}^{\phi(k)} P(\phi_{0}) d\phi_{0}}\\
&=& \frac{\int_{k}^{\infty} \exp{\frac{3}{8V_{0}\left( 1 - 2^{-2k'} \right)^{2}}} k'^{2} 2^{-k'} dk'}{\int_{k}^{\infty} \exp{\frac{3}{8V_{0}\left( 1 - 2^{-2k'} \right)^{2}}} 2^{-k'} dk'}.
\end{eqnarray}
From $F_{1}$ and $F_{2}$, we define (Right of Figure~\ref{fig:F1})
\begin{eqnarray}
F_{3}(k,V_{0}) \equiv \sqrt{F_{2}(k) - F^{2}_{1}(k)}.
\end{eqnarray}
Again, for some special limiting cases, we can calculate further. If $V_{0} \sim 1$,
\begin{eqnarray}
F_{3}(k,V_{0}) \simeq 1.44
\end{eqnarray}
for $k > 0.5$. On the other hand, if $V_{0} \ll 1$, then
\begin{eqnarray}
F_{3} (k,V_{0}) \simeq 0.
\end{eqnarray}

\begin{figure}
\begin{center}
\includegraphics[scale=0.4]{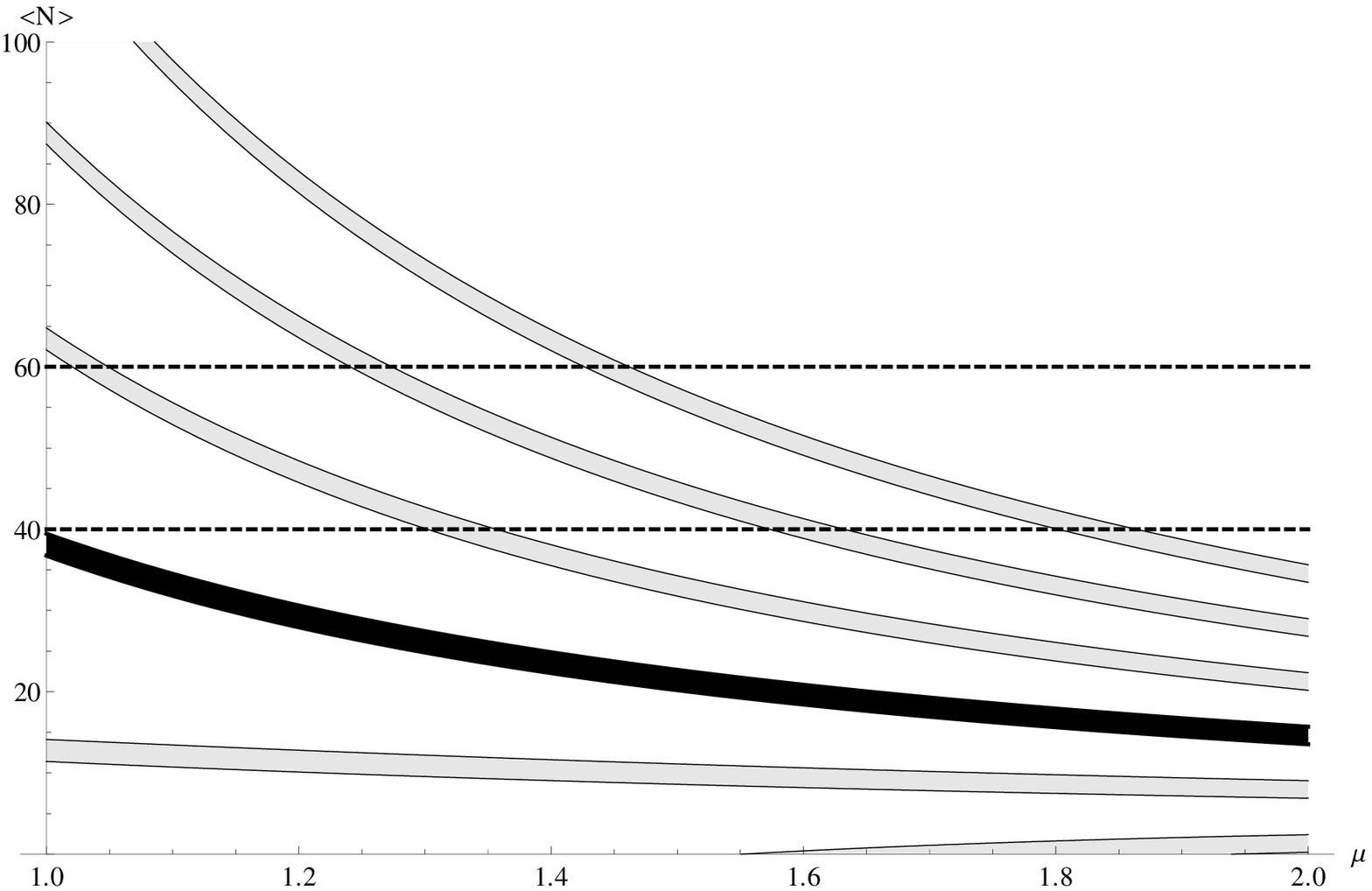}
\includegraphics[scale=0.405]{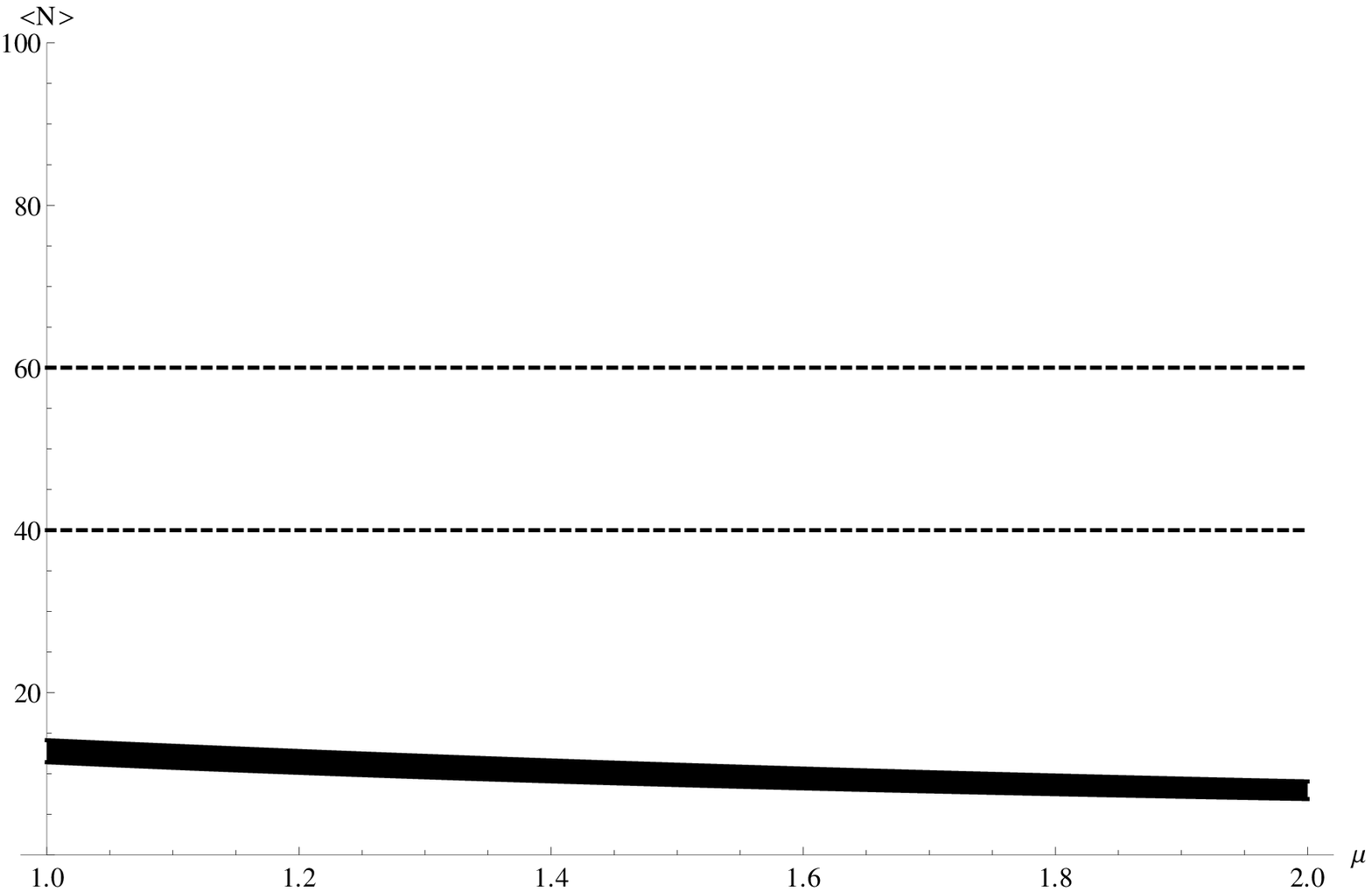}
\caption{\label{fig:expectationvalue}Left: For $V_{0}\simeq 1$. Thick black region is $\langle\mathcal{N}\rangle$ as a function of $\mu$. Gray regions are $\langle\mathcal{N}\rangle \pm \Delta \mathcal{N}, \pm 2\Delta \mathcal{N}, \pm3\Delta \mathcal{N}$, respectively. Right: As $V_{0}$ significantly decreases, $\langle\mathcal{N}\rangle$ approaches to the black curve.}
\end{center}
\end{figure}

Up to now, all quantity depends on the domain of integration $dk'$, where $k < k' < \infty$; we choose $k = k_{\mathrm{min}}$ and it is related to the cutoff and hence depends on $\mu$. Now we calculate the expectation value of the number of $e$-foldings:
\begin{eqnarray}
\langle\mathcal{N}\rangle &=& \frac{\int \mathcal{N} P(\phi)d\phi}{\int P(\phi) d\phi}\\
&=& C \mu^{-p} F_{1}(k_{\mathrm{min}},V_{0}).
\end{eqnarray}
To calculate the dispersion, we first calculate such that
\begin{eqnarray}
\langle\mathcal{N}^{2}\rangle &=& \frac{\int \mathcal{N}^{2} P(\phi)d\phi}{\int P(\phi) d\phi}\\
&=& C^{2} \mu^{-2p} F_{2}(k_{\mathrm{min}},V_{0}).
\end{eqnarray}
Then, the standard deviation becomes
\begin{eqnarray}
\Delta \mathcal{N} &=& \sqrt{\langle\mathcal{N}^{2}\rangle - \langle\mathcal{N}\rangle^{2}}\\
&=& C \mu^{-p} F_{3}(k_{\mathrm{min}},V_{0}).
\end{eqnarray}
We can calculate for some limiting cases:
\begin{description}
\item[- $V_{0} \simeq 1$ limit:]
For $k_{\mathrm{min}} > 0.5$, approximately,
\begin{eqnarray}
\langle\mathcal{N}\rangle &\simeq& 17.59 \times \mu^{-1.93} \times k_{\mathrm{min}}[\mu] + 17.59 \times 1.44 \times \mu^{-1.93},\\
\Delta \mathcal{N} &\simeq& 25.33 \times \mu^{-1.93}.
\end{eqnarray}
We use $k_{\mathrm{min}}=k_{\mathrm{c}}$ for the lowest bound estimation and $k_{\mathrm{min}}=k_{\mathrm{M}}$ for the largest bound estimation. Finally, the results are summarized in left of Figure~\ref{fig:expectationvalue}.
\item[- $V_{0} \ll 1$ limit:]
\begin{eqnarray}
\langle\mathcal{N}\rangle &\simeq& 17.59 \times \mu^{-1.93} \times k_{\mathrm{min}}[\mu],\\
\Delta \mathcal{N} &\simeq& 0.
\end{eqnarray}
Hence, the standard deviation is very small. If $\mu \simeq 1$, then the expected $e$-folding is approximately $11 \lesssim \langle\mathcal{N}\rangle \lesssim 15$. This is summarized in right of Figure~\ref{fig:expectationvalue}.
\end{description}
Note that as long as $\mu \gtrsim 1$, the standard deviation is reasonably small and hence these expectation values are meaningful.

In conclusion, when we assume the hill-top potential, if the inflation was began at the high energy scale $V_{0} \gtrsim 10^{-2}$, then we reasonably expect around $30$ $e$-foldings. If the inflation was began at the low energy scale $V_{0} \ll 1$ around the hill-top, then it will only contribute to $\sim 10$ $e$-foldings.

\begin{figure}
\begin{center}
\includegraphics[scale=0.5]{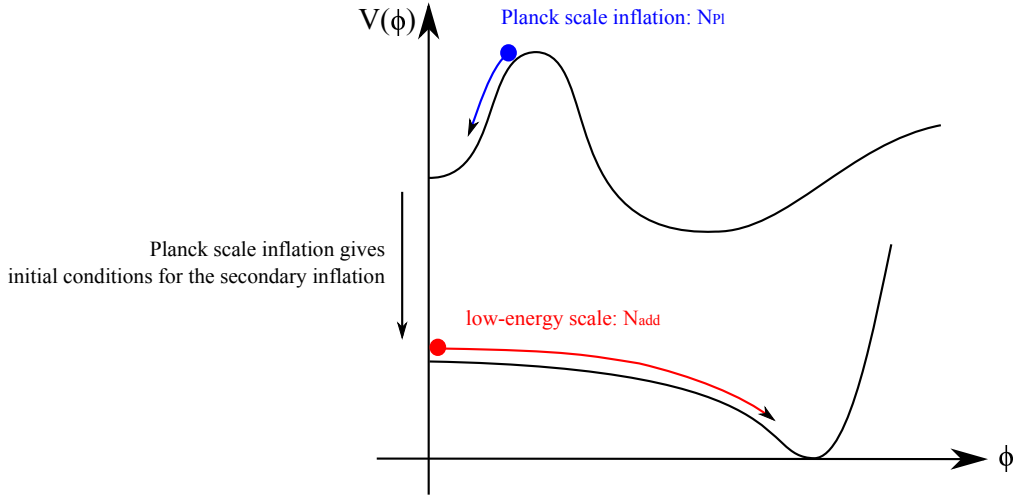}
\caption{\label{fig:potential3}Conceptual scenario of the Planck scale inflation.}
\end{center}
\end{figure}

\paragraph{Possible scenarios and interpretations}

We illustrate possible scenarios to build a suitable model according to (P1) (Figure~\ref{fig:potential3}).

\begin{enumerate}
\item At the beginning of the universe, the energy scale was close to the Planck regime $V_{0} \simeq 1$ and perhaps string theory contributions were important. The no-boundary proposal cause the first pre-inflation. There was a natural hill-top potential $\mu \gtrsim 1$. Or, there are a number of fields with various hill-tops ($m_{i} \sim m_{0}$) and their effective mass around the tops was $\mu \gtrsim 1$, where $\mu^{2} \simeq m^{2}_{0}/\sum V_{0,i}$. This made almost $\mathcal{N}_{\mathrm{Pl}} \simeq 30$ $e$-foldings. This stage can explain the dilaton/moduli stabilization \cite{Hwang:2011mp,Hwang:2012zj}. 

\item Next -- via the symmetry breaking, decreasing of the temperature, or any other effects -- around a certain energy scale $V_{0} \ll 1$, the second hill-top inflation began again (of course, in the Lorentzian signatures). Hence, \textit{the pre-inflationary stage gives a good initial condition for the secondary inflation; and, we can interpret that the no-boundary proposal prefers such a pre-inflationary era}. This adds $\mathcal{N}_{\mathrm{add}} > 10$ or larger number of $e$-foldings.

We can check this possibility using a simple calculation. For a small field inflation occurring for $\phi \ll \phi_\mathrm{m}$, the expansion rate is given by
\beq
H_I \simeq \sqrt{\frac{\lambda}{12}} \frac{\phi_\mathrm{m}^2}{M_{\rm P}}.
\eeq
Assuming $\phi$ is homogeneous and solving the equation of motion of $\phi$,
\beq
0 = \ddot{\phi} - 3 H \dot{\phi} + V',
\eeq
where $'$ denotes the derivative with respect to $\phi$, one finds an approximate solution
\beq
\phi(t) = \phi_i e^{\alpha_\phi H \left( t - t_i \right)}
\eeq
where $\phi_i$ is the field value at a time $t_i$ and
\beq
\alpha_\phi = \frac{3}{2} \left( \sqrt{1 + \frac{4 m_\phi^2}{9 H^2}} - 1 \right)
\eeq
with
\beq
m_\phi^2 \simeq \lambda \phi_\mathrm{m}^2 \simeq 12 H_I^2 \left( \frac{M_{\rm P}}{\phi_\mathrm{m}} \right)^2
\eeq
being the mass of the inflaton.
The number of $e$-foldings is given by
\beq
\mathcal{N}_{\mathrm{add}} \simeq \frac{1}{\alpha_\phi} \log \left( \frac{\phi_\mathrm{m}}{\phi_*} \right)
\eeq
where $\phi_*$ is the field value when a cosmological scale of interest exit horizon. For $\phi_\mathrm{m} = M_{\rm P}$ and $\phi_* \gtrsim 10^{-13} M_{\rm P}$ as an example, we find $\mathcal{N}_{\mathrm{add}} \lesssim 13$. In this case, the previous stage gives the initial condition for the secondary inflation; hence, the scalar field could begin from the top of the potential. This field will follow usual rules of the quantum field theory in de Sitter space. This stage can be repeated some times.

\item Perhaps, after the inflation ends, the curvaton is now dominated (if necessary). This makes fluctuations and other observables. 

\item Perhaps, there may be further inflation (e.g., thermal inflation) that is not harmful to curvature perturbations and only adds order $\mathcal{N}_{\mathrm{fin}} \simeq 10$ $e$-foldings.
\end{enumerate}
Therefore, the final $e$-folding becomes $\mathcal{N} = \mathcal{N}_{\mathrm{Pl}} + \mathcal{N}_{\mathrm{add}} + \mathcal{N}_{\mathrm{fin}}$. If $\mathcal{N}_{\mathrm{add}} + \mathcal{N}_{\mathrm{fin}} > \mathcal{N}_{*}$, then we cannot see the effects of the Planck scale inflation. On the other hand, if $\mathcal{N}_{\mathrm{add}} + \mathcal{N}_{\mathrm{fin}} < \mathcal{N}_{*}$, then there is some hope to see the effects.

The details of phenomenology are highly model dependent. However, there can be some observational effects as long as $\mathcal{N}_{\mathrm{add}} + \mathcal{N}_{\mathrm{fin}} < \mathcal{N}_{*}$. One educative example is \cite{Namjoo:2012xs}. In this paper, the authors study when there are multiple slow-roll inflationary stages. The detailed observational consequences on the density perturbations depend on the details of the transition process. One typical outcome of this scenario is that perhaps it predicts an overproduction of primordial black holes than the single field inflation models. Of course, to say more details, we should fix the model for the (sub-) Planck scale inflation and the secondary inflation, as well as the transition process; and perhaps, we need to include some effects of curvatons. In this sense, the results of \cite{Namjoo:2012xs} should be extended for our cases and it goes beyond the scope of the present paper. However, what we can safely say is that (1) the no-boundary proposal is more compatible with the high energy scale pre-inflation model and (2) this can remain some observational imprints, in principle\footnote{It is worthwhile to notice on the recent tension between Planck data \cite{Ade:2013uln} and the BICEP2 data \cite{bicep2}. Some of pre-inflationary effect can explain the tension between two data, although it needs further clarification.}.

\subsubsection{\label{sec:bo}Bounded and unbounded slow-roll hill-top inflation}

In this subsection, we consider the case (P2) so that the potential in itself allows small $\Delta$ or large $\Xi$. First we test for large field slow-roll inflation model and we comment that it cannot be a good model in terms of typicality. However, by choosing a potential, we can obtain proper models.

\paragraph{Bounded hill-top inflation: large field inflation}

\begin{figure}
\begin{center}
\includegraphics[scale=0.25]{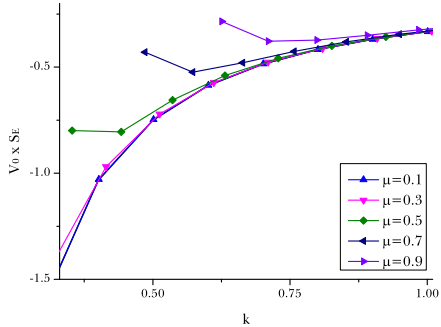}
\caption{\label{fig:probability_slow}The Euclidean action $\tilde{S}_{\mathrm{E}}$ as a function of $k$ by varying $\mu$, in the slow-roll limit. There is too much hierarchy for $\Delta$ and hence not compatible with the no-boundary proposal.}
\end{center}
\end{figure}

In Figure~\ref{fig:probability_slow}, we consider the no-boundary proposal for slow-roll limit $\mu < 1$. Then the action difference $\Delta$ increases: in Figure~\ref{fig:probability}, it was order of $0.1$, while in Figure~\ref{fig:probability_slow}, it is order of $1$. Therefore, for this naive potential, it is not easy to explain the preference of large $e$-foldings.

\paragraph{Bounded hill-top inflation: small field inflation}

To construct a small field inflation model, we use the following form:
\begin{eqnarray}
V(\phi) &=& V_{0} V_{\mathrm{CW}}(\phi) + \Lambda,\\
V_{\mathrm{CW}}(\phi) &=& \left[ \left( \frac{\lambda}{6} - \frac{9e^{4}}{32 \pi^{2}} \right) + \frac{3 e^{4}}{16 \pi^{2}} \log \frac{2 e^{2} \phi^{2}}{M^{2}} \right] \phi^{4+2n},
\end{eqnarray}
where $\Lambda$ is the cosmological constant, $V_{0}$ is the overall constant that can be chosen freely by the rescaling, $e$ is the gauge coupling, $\lambda$ is the self coupling of $\phi$, $M$ is the cutoff scale, and $n$ is the parameter for the higher order potential, while $n=0$ is the case for the Coleman-Weinberg potential \cite{Coleman:1973jx}.

\begin{figure}
\begin{center}
\includegraphics[scale=0.5]{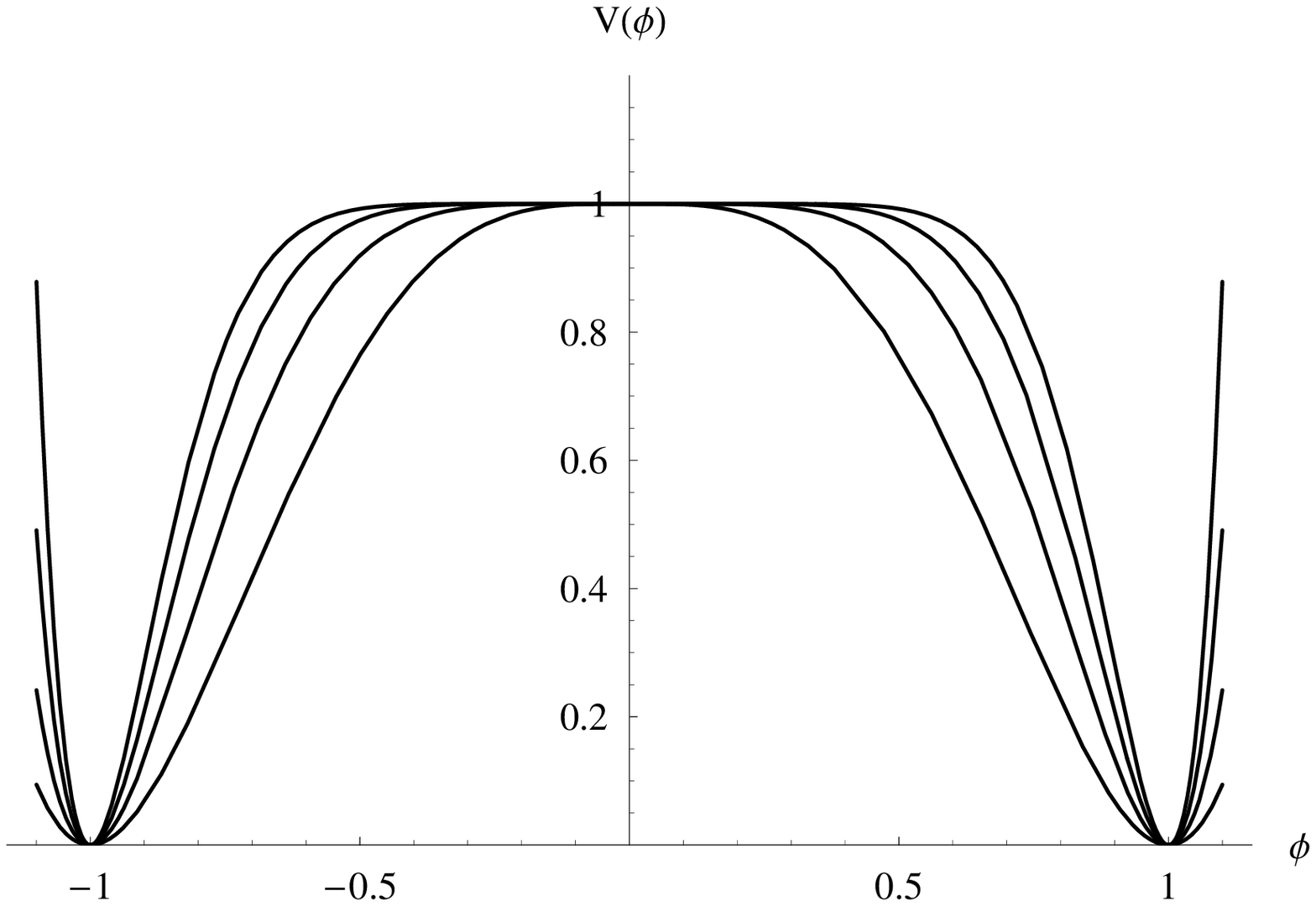}
\caption{\label{fig:CW}$V(\phi)$ for $n=0, 1, 2, 3.$}
\includegraphics[scale=0.25]{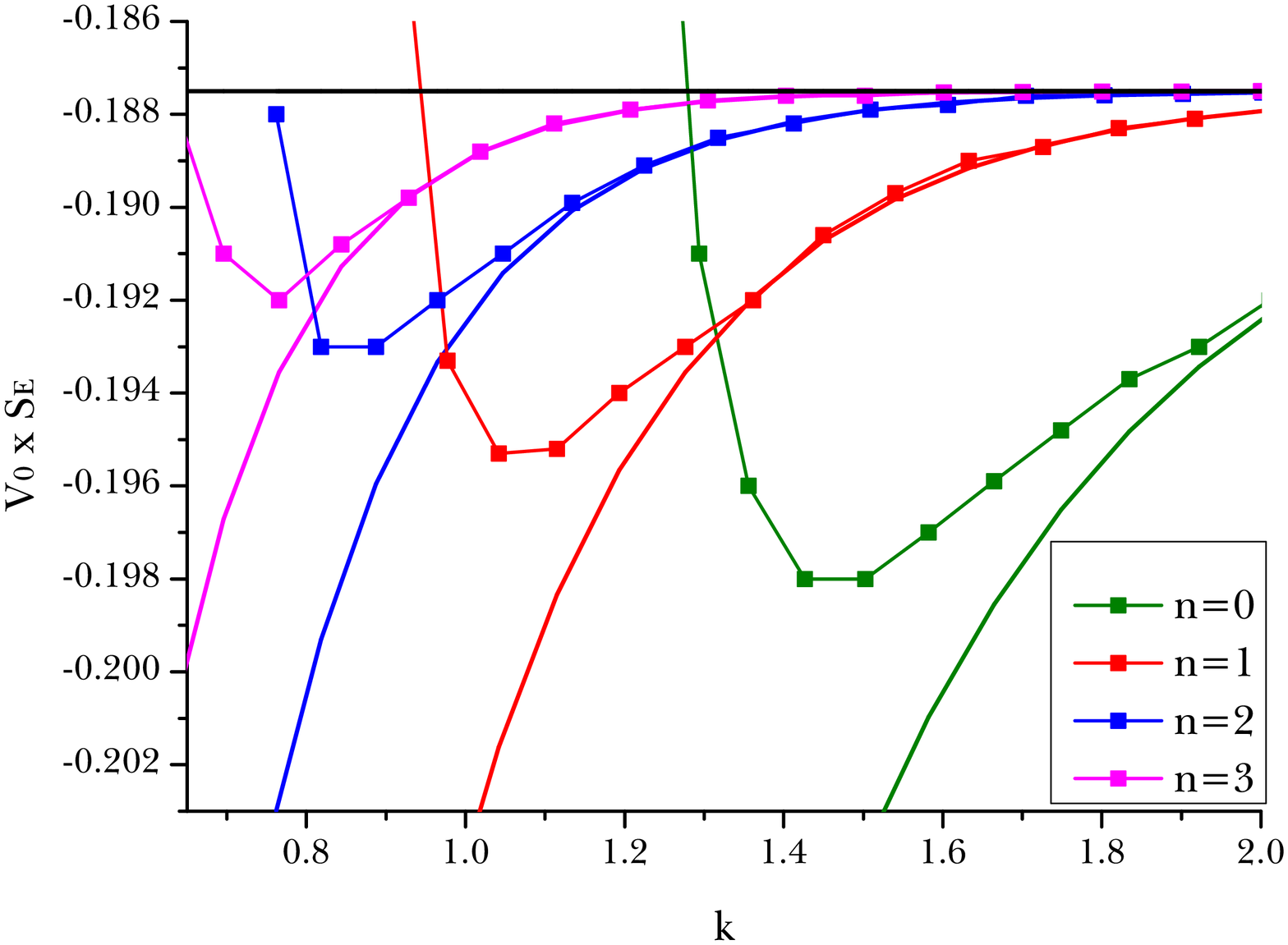}
\includegraphics[scale=0.25]{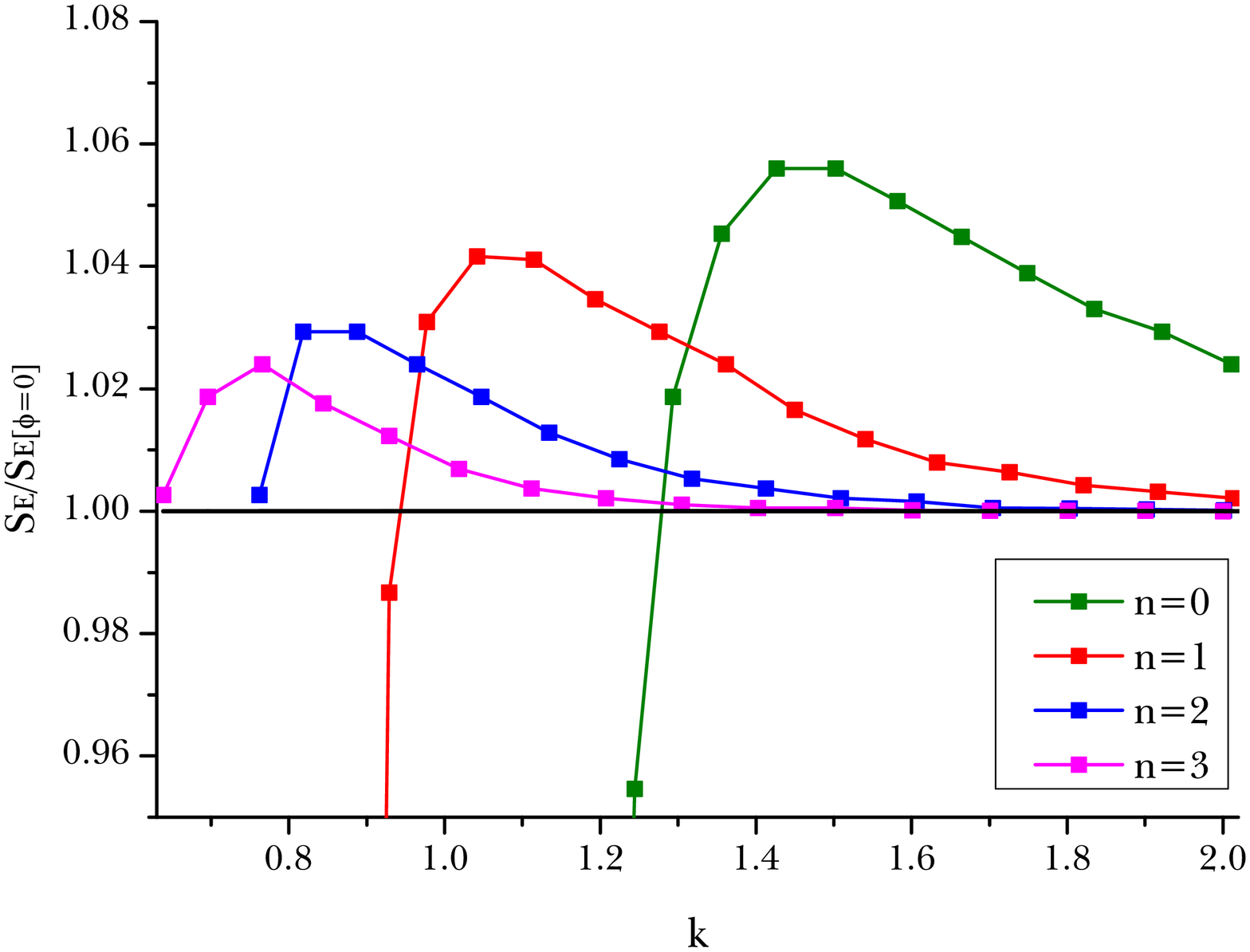}
\caption{\label{fig:CW_result}Left: $\tilde{S}_{\mathrm{E}}$ for $n=0, 1, 2, 3.$ Thick curves are for analytic fitting by assuming slow-roll; thick black curve is the case for the Hawking-Moss instanton at the local maximum. Right: $S_{\mathrm{E}}/S_{\mathrm{E}}[\phi=0]$ for $n=0, 1, 2, 3.$}
\includegraphics[scale=0.25]{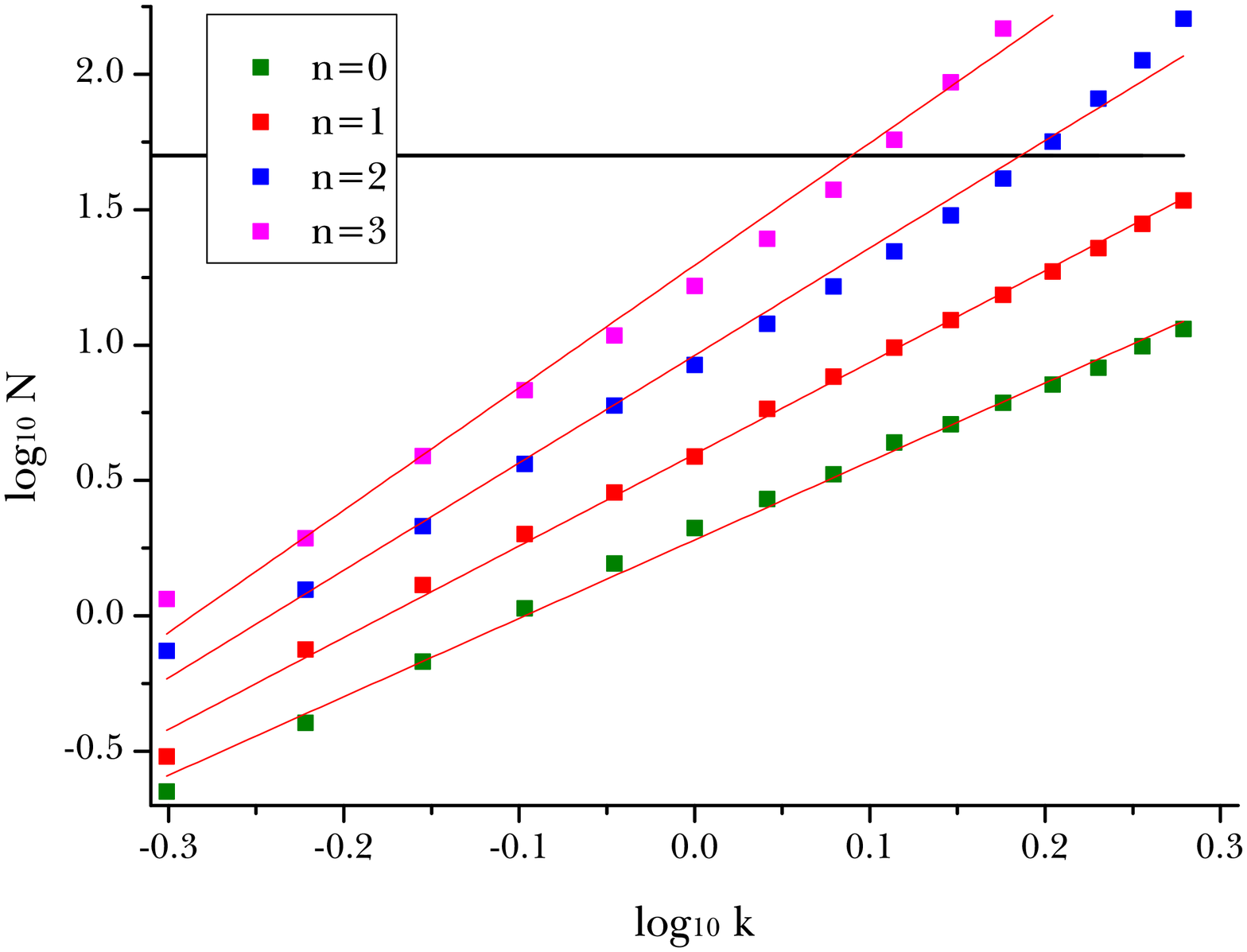}
\caption{\label{fig:efoldings_CW}The number of $e$-foldings is estimated for $n=0, 1, 2, 3.$ Black thick horizontal line is $\mathcal{N}=\mathcal{N}_{*}$.}
\end{center}
\end{figure}

Here, we choose $M$ to make $V'(1) = 0$:
\begin{eqnarray}
M = \sqrt{2}e \exp \left[ \frac{1}{4n+2} + \frac{4\pi^{2} \lambda}{9 e^{4}} - \frac{3}{4} \right].
\end{eqnarray}
In addition, it is natural to choose $\Lambda$ to satisfy $V(1) = 0$. Furthermore, by re-scaling, we can make $V(0)=1$, without loss of generality. For convenience, we choose $e=0.1$ and $\lambda=0.00001$. Note that the potential shape is not so sensitive with the parameters, and hence we can choose them sufficiently small. Figure~\ref{fig:CW} summarize for various $n$.

\begin{figure}
\begin{center}
\includegraphics[scale=0.35]{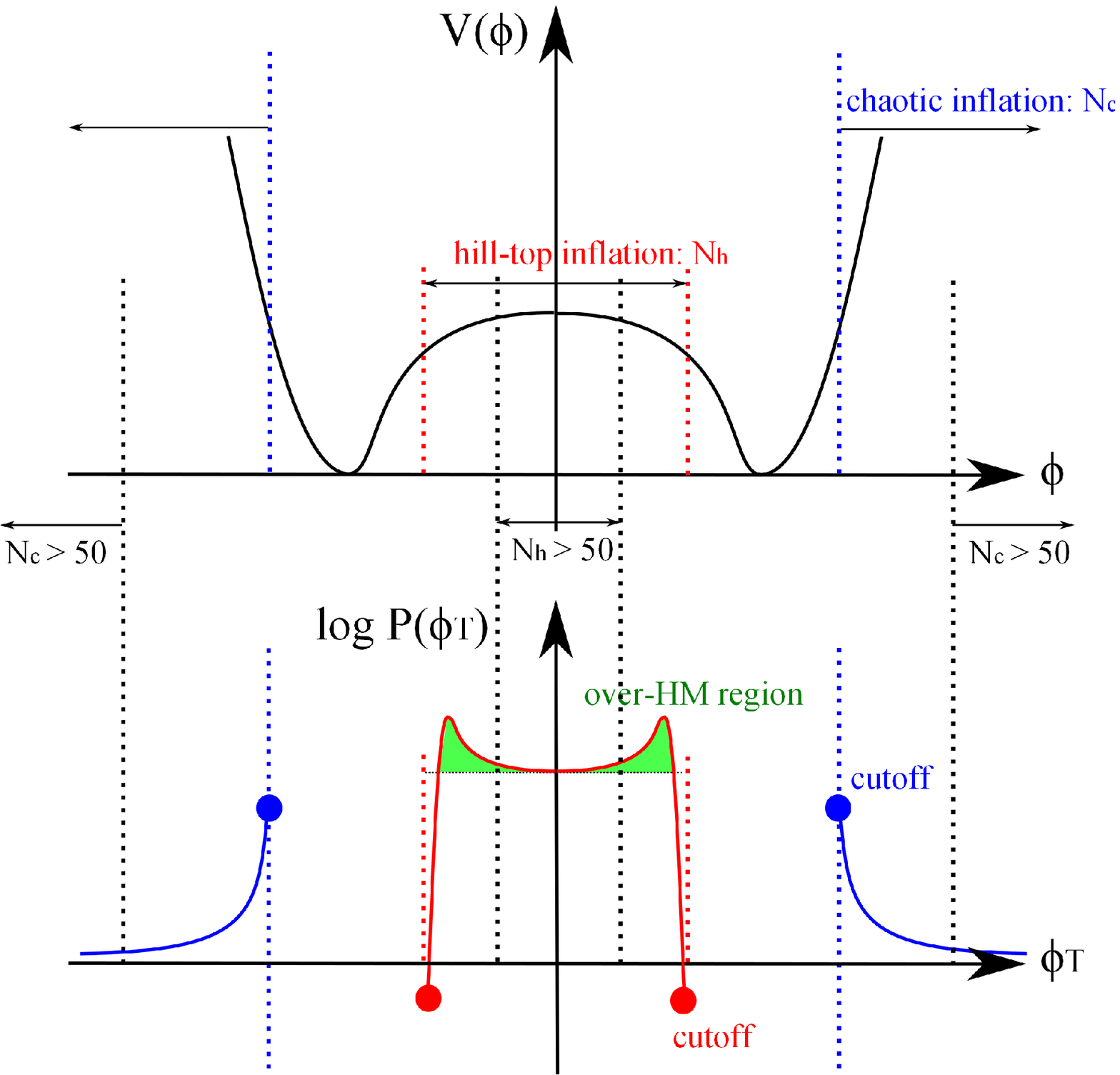}
\includegraphics[scale=0.35]{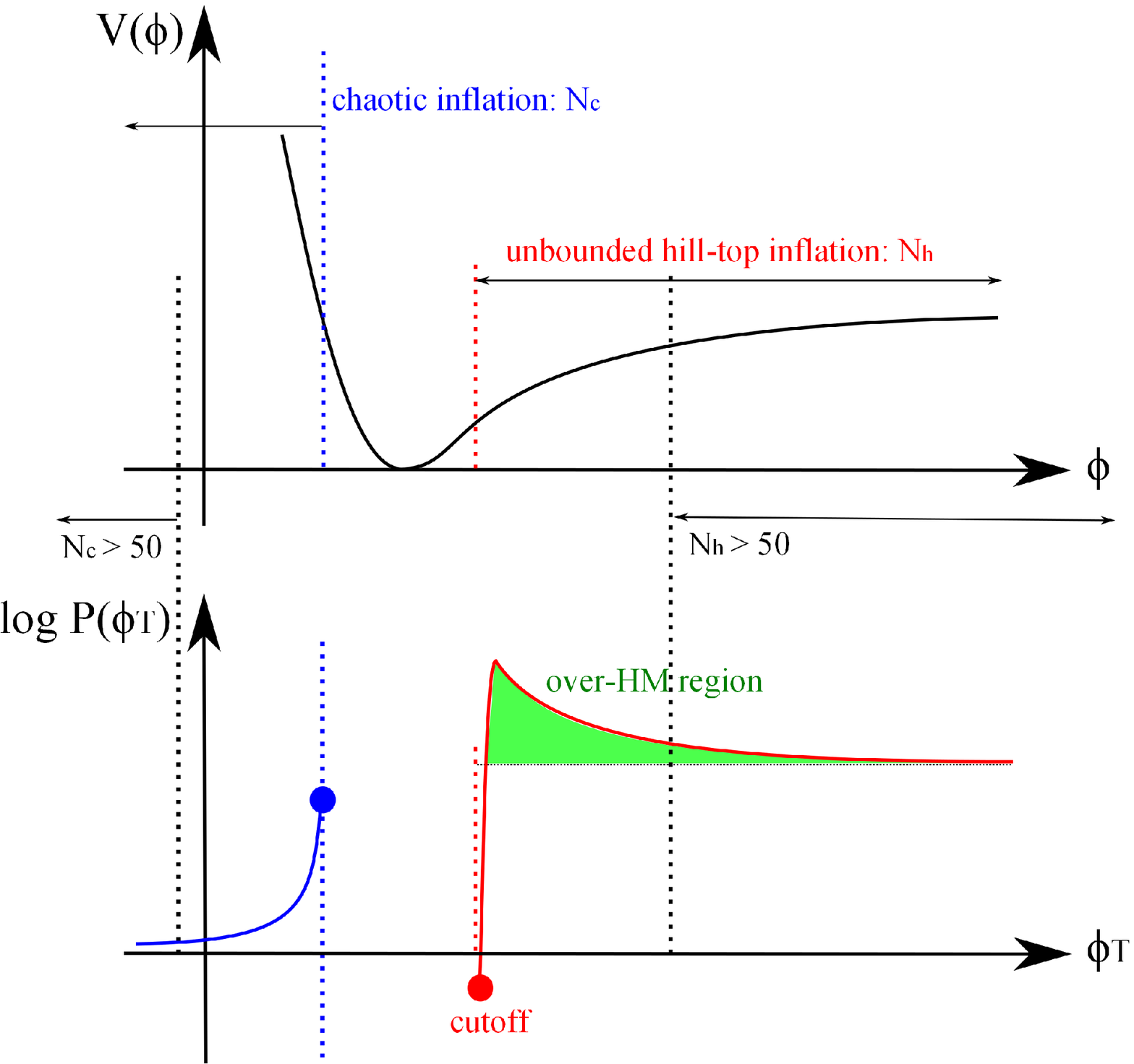}
\caption{\label{fig:potential}Left: Conceptual figure of a hill-top potential and the no-boundary proposal. The chaotic inflation cannot give sufficient $e$-foldings, because exponentially disfavored. Sufficient $e$-foldings of the hill-top region can be reasonable as long as the region with higher probability than the Hawking-Moss instanton of the hill-top (over-HM region; green colored region) is sufficiently small. Right: For some scalar-tensor type models, there is a non-compact field direction so that it allows sufficient field space for sufficient $e$-foldings.}
\end{center}
\end{figure}

Figure~\ref{fig:CW_result} denote the Euclidean probability. For $n=1$ case, already the action difference is around $1 \%$. Since the potential is sufficiently flat on the top while it is quite steep around the local minimum, the hierarchy between the highest probability region and the lowest probability region gradually decreases as $n$ increases. In Figure~\ref{fig:efoldings_CW}, the number of $e$-foldings is estimated: here,
\begin{eqnarray}
\log_{10} \mathcal{N} \simeq D \log_{10} k + E
\end{eqnarray}
is a good estimation. As we vary $n=0, 1, 2, 3$, the gradients $D$ are $2.896$, $3.387$, $3.967$, $4.518$; intercepts $E$ are $0.280$, $0.597$, $0.961$, $1.294$, respectively. The largest probability points $k_{\mathrm{M}}$ are $1.356$, $0.977$, $0.888$, $0.765$, respectively; then, $e$-foldings for each $k_{\mathrm{M}}$ are $4.6$, $3.7$, $5,7$, $5.9$, respectively. Therefore, around the largest probability point, we cannot get a sufficient $e$-foldings. To obtain a reasonable typicality, we should increase $n$ so that we should decrease $\Delta$ sufficiently (Left of Figure~\ref{fig:potential}).

One interesting point is that this model prefers the hill-top inflation rather than the chaotic inflation, as long as the potential around the local minimum steeper and steeper (left of Figure~\ref{fig:potential}). Therefore, this can be an answer for the doubt C in Section~\ref{sec:dob}.

\paragraph{Unbounded hill-top inflation: $R^{2}$/scalar-tensor inflation}

The Starobinski model of $R^{2}$ inflation \cite{Starobinsky:1980te} is
\begin{eqnarray}
S = \int \sqrt{-g} dx^{4} \frac{1}{16\pi} \left[R + \frac{1}{2} m^{2} R^{2} + ... \right].
\end{eqnarray}
After the field re-definition, it becomes a general scalar-tensor gravity model:
\begin{eqnarray}
S = \int \sqrt{-g} dx^{4} \left[ \frac{1}{16\pi} \Phi R - \frac{\omega}{\Phi} \left(\nabla \Phi\right)^{2} - \frac{1}{2m^{2}} \left( \Phi - 1 \right)^{2} + ... \right]
\end{eqnarray}
with $\omega = 0$. After the conformal transformation, the potential in the Einstein frame becomes
\begin{eqnarray}
V(\Phi(\phi)) = \frac{1}{2m^{2}}\frac{\left( \Phi - 1 \right)^{2}}{\Phi^{2}},
\end{eqnarray}
where $\phi$ is the canonical scalar field in the Einstein frame. This is also the same for the non-minimally coupled inflation model:
\begin{eqnarray}
S = \int \sqrt{-g} dx^{4} \left[ \frac{1}{16\pi} \phi^{2} R - \frac{1}{2} \left(\nabla \phi\right)^{2} - \frac{1}{2} m^{2} \left(\phi - \phi_{\mathrm{m}}\right)^{2} + ... \right].
\end{eqnarray}

For these models, whatever the Euclidean action (the authors calculated in \cite{Hwang:2011mp}) is,
\begin{eqnarray}
\Xi \equiv \log \frac{\phi_{\mathcal{N}_{*}} - \phi_{\mathrm{top}}}{\phi_{\mathrm{M}}-\phi_{\mathcal{N}_{*}}}
\end{eqnarray}
should diverge, since the field space is unbounded; if we choose $\phi_{\mathrm{top}}=0$, then this implies that $\phi_{\mathcal{N}_{*}}$ diverges while $\phi_{\mathrm{M}}-\phi_{\mathcal{N}_{*}}$ is bounded. Apparently, these models are good in terms of the no-boundary proposal (Right of Figure~\ref{fig:potential}). However, it assumes that the model is still valid for infinitely large $\Phi$. If it is not, then these may not be a good model.

\subsubsection{\label{sec:ch}Chaotic inflation and multi-field inflation}

In this subsection, we consider the possibility (P3). First we summarize the no-boundary proposal for simple chaotic inflation. This can be generalized for a multi-field inflation model. The multi-field configuration can contribute as a new ingredient for the probability and the typicality.

\paragraph{Euclidean probability for chaotic inflation} First let us consider the potential
\begin{equation}
V(\phi) = \lambda \phi^{n}.
\end{equation}
To estimate the no-boundary proposal for general chaotic inflation models, we use the approximation method used in \cite{Lyons:1992ua}.
The following solutions are approximate solutions of the equations of motion:
\begin{eqnarray}
\phi \simeq \phi_{0} + i \beta \tau, \;\;\;\; \rho \simeq i \frac{3\beta}{n\lambda\phi_{0}^{n-1}} \exp \left[ -\frac{\lambda}{3\beta^{2}} \phi_{0}^{n} \left( \left(1 + i \frac{\beta}{\phi_{0}} \tau \right)^{n}-1 \right) \right],
\end{eqnarray}
in which the scalar field $\phi$ slowly rolls.
If the scalar field rolls more slowly, then we can further approximate
\begin{eqnarray}
\label{approx2}\phi \simeq \phi_{0}, \;\;\;\; \rho \simeq i \frac{3\beta}{n\lambda\phi_{0}^{n-1}} \exp\left[ -i \frac{n\lambda}{3\beta} \phi_{0}^{n-1} \tau\right].
\end{eqnarray}
Hence, the real part of $\rho$ becomes the sine function:
\begin{eqnarray}
\rho^{\mathrm{re}} \simeq \frac{3\beta}{n\lambda\phi_{0}^{n-1}} \sin \left[ \frac{n\lambda}{3\beta} \phi_{0}^{n-1} \tau\right].
\end{eqnarray}
Note that $n\lambda \phi_{0}^{n-1}/3\beta = H$, where $H$ is the Hubble parameter of the slow-roll inflation and we should choose turning point by $X = \pi/2H$. Therefore, we can think that it follows the behavior of the Hawking-Moss instanton \cite{Hawking:1981fz} in the slow-roll limit.
The Euclidean action and the result is
\begin{eqnarray}
S_{\mathrm{E}} = 4\pi^{2} \int_{0}^{X} \left( \rho^{3} V - \frac{3}{8 \pi} \rho \right) d \tau \simeq - \frac{3}{16 V(\phi_{0})}.
\end{eqnarray}

To check the typicality, we approximately calculate that
\begin{eqnarray}
\Delta \simeq \left|\frac{3}{16 V(\phi_{\mathcal{N}_{*}})} - \frac{3}{16 V(\phi_{\mathrm{cutoff}})}\right|
\end{eqnarray}
and $V(\phi_{\mathrm{cutoff}}) \simeq 1$ and $V(\phi_{\mathcal{N}_{*}}) \gtrsim 1$. Therefore, $\Delta \gtrsim \lambda^{-1} \times \mathcal{O}(1)$. Note that to explain the observational constraints, $\lambda \ll 1$ is required. Therefore, the naive chaotic inflation scenario cannot be a good hypothesis in terms of the no-boundary proposal.

\paragraph{Generalization to multi-field inflation}

On the other hand, if we consider assisted inflation with many numbers of fields \cite{DKMW}, then there may be a good way to enhance the large $e$-foldings. In this subsection, we summarize the results of \cite{Hwang:2012bd}.

Because of the technical limitation on the use of the no-boundary proposal, we will approximate the model as follows: let us consider that there are $N$ number of fields $\phi_{i}$ with the same mass $m$ and each field has the potential
\begin{eqnarray}
V(\phi_{1},\phi_{2}, ... , \phi_{N}) = \frac{1}{2} m^{2} \sum_{i=1}^{N} \phi_{i}^{2}.
\end{eqnarray}
Since all masses are the same, we can define the modulus of the field:
\begin{eqnarray}
\vec{\phi} = (\phi_{1}, \phi_{2}, ... , \phi_{N})
\end{eqnarray}
like a vector. Since the no-boundary wave function effectively makes the velocity at $\tau = 0$ to be zero, the angular velocity of $\vec{\phi}$ is always zero. Hence, this model is effectively the same as a single quadratic scalar field $\Phi = |\vec{\phi}|$, with the same mass $m^{2}$.

\begin{figure}
\begin{center}
\includegraphics[scale=0.4]{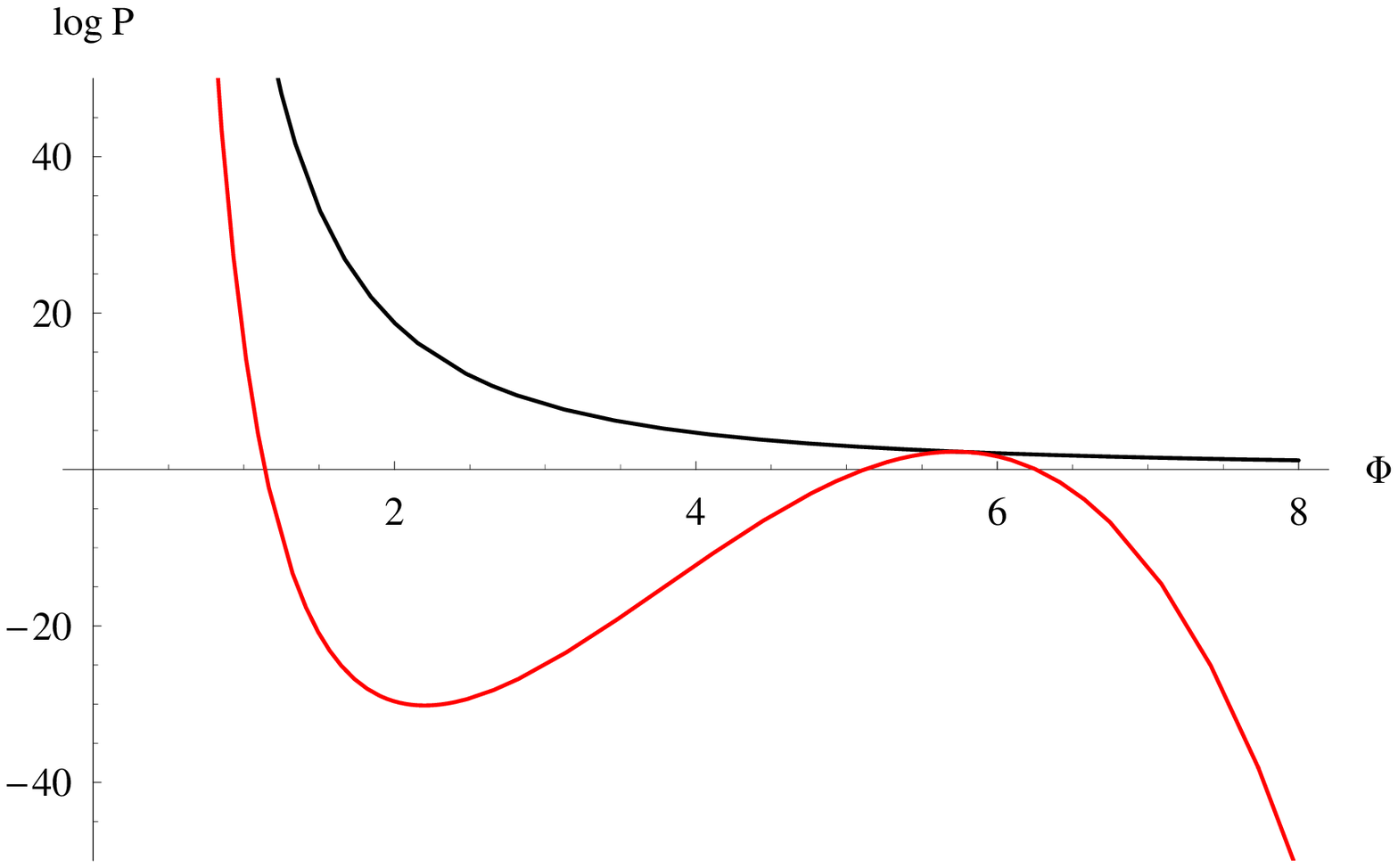}
\includegraphics[scale=0.4]{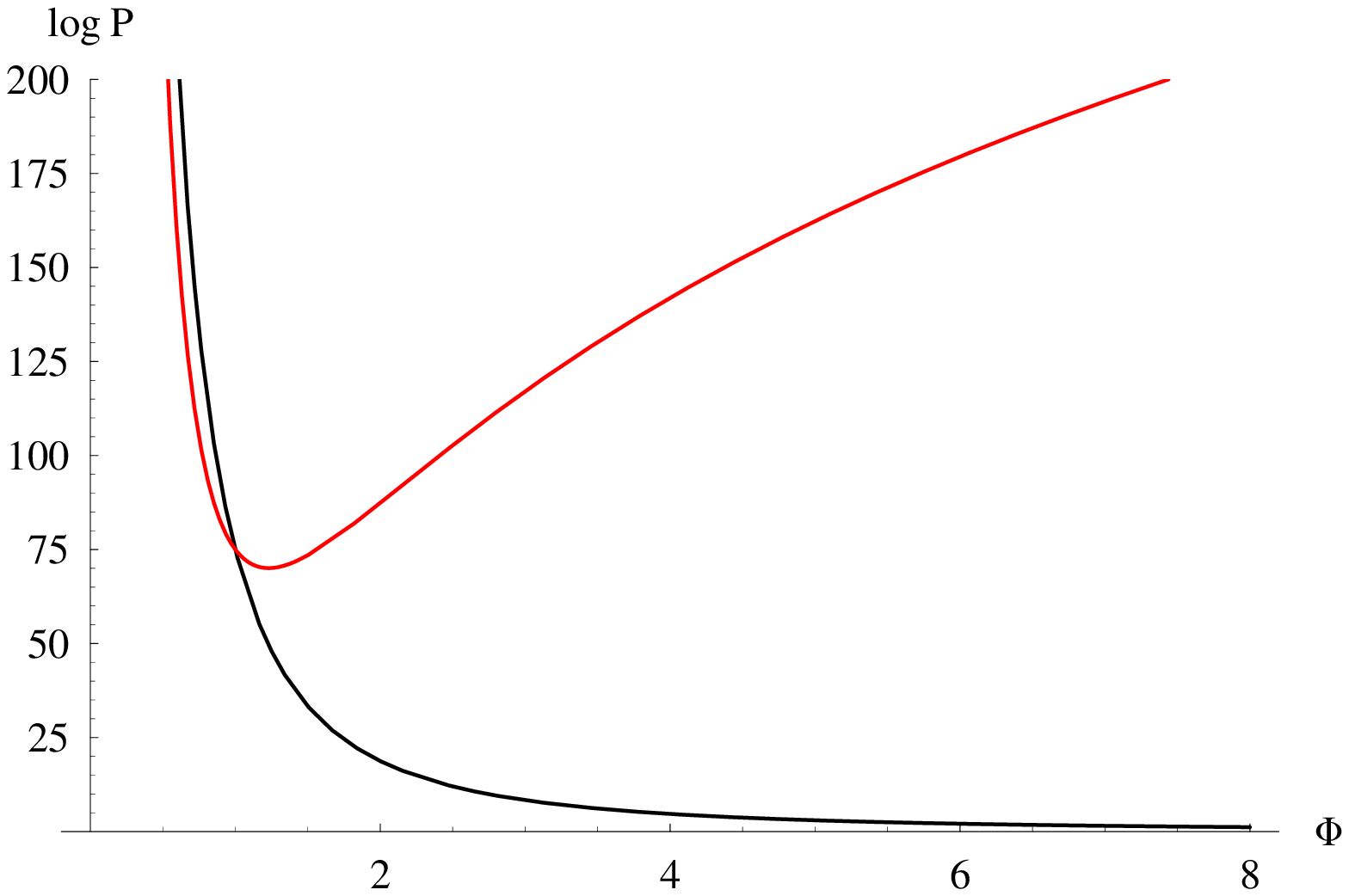}
\caption{\label{fig:nf}Left: red curve is $\log P$ by assuming usual cutoff $0 \leq \left|\phi_{i}\right| < 1$ (Equation~(\ref{eq:nf})). Right: red curve is $\log P$ without assuming the cutoff, and only regarding the phase volume with $N$ (Equation~(\ref{eq:nf2})). In these plots, we assumed $N=m^{-2}=100$ and they show that two factors -- Euclidean action and degeneracy factor -- compete each other. Black curves are the no-boundary wave function without degeneracy effects.}
\end{center}
\end{figure}

However, since we introduce many numbers of fields, the phase volume $\Pi^{N}_{i=1} d\phi_{i}$ of a large field modulus becomes greater than that of the minimum. In addition, as usual multi-field inflation models, it is not unreasonable to bound field values less than the Planck scale: $\phi_{i} \lesssim 1$. Then we can precisely add the effects of the phase volume, or equivalently the degeneracy of field spaces, as follows \cite{Hwang:2012bd}:
\begin{eqnarray}\label{eq:nf}
dP\left[ \Phi \right] \simeq \exp \left[ \frac{3}{4m^{2} \Phi^{2}} - \frac{\left( \Phi^{2} - N/3 \right)^{2}}{8N/45} \right] d\Phi.
\end{eqnarray}
Therefore, around the cutoff $\Phi \simeq 1$, the probability is approximately $\log P \sim m^{-2} - N$. On the other hand, around the mean value $\Phi \sim N/3$, the probability is approximately $\log P \sim m^{-2}N^{-1}$. To make two values be similar orders, we require $N \simeq m^{-2}$. Left of Figure~\ref{fig:nf} shows that the large $N$ effects can be competitive with the peak of the left side cutoff.

This fact does not crucially depend on the details of the cutoff. If we do not insert the cutoffs, then we have to include the effects of the phase volume. The phase volume is
\begin{eqnarray}
\Pi^{N}_{i=1} d\phi_{i} = \Phi^{N-1} d\Phi d\Omega_{N-1}
\end{eqnarray}
and hence
\begin{eqnarray}\label{eq:nf2}
dP\left[ \Phi \right] \simeq \exp \left[ \frac{3}{4m^{2} \Phi^{2}} + \left(N-1\right) \log \Phi \right] d\Phi.
\end{eqnarray}
The exponential has two maxima: $\Phi_{\mathrm{cutoff}}$ and $\Phi_{\mathrm{max}}$, where $\Phi_{\mathrm{max}}$ should be a kind of cutoff in the field space. If $N \simeq m^{-2}$, then $\Phi_{\mathrm{cutoff}}$ and $\Phi_{\mathrm{max}}$ can be a similar order, and hence, this enhances probability of large $e$-foldings sufficiently: right of Figure~\ref{fig:nf}.

Of course, two results are different, since we applied different assumptions to cut the field space. However, two approaches give the same conclusion: \textit{as long as we have $N \simeq m^{-2}$ number of fields, then large $e$-foldings can be comparable with the small $e$-foldings.}
Since the $e$-folding is
\begin{equation}
\mathcal{N} = 2\pi \Phi^{2},
\end{equation}
as long as $N \gtrsim 30$, the $e$-folding at the mean value of the Gaussian peak in Equation~(\ref{eq:nf}) is greater than $\mathcal{N}_{*}$, and hence, it can be a good hypothesis.

Of course, this model in itself is not quite compatible with the Planck data. However, if we include the contribution of the curvaton, we can make this model compatible with observational constraints. If we can choose $m^{2}\simeq {10^{-2}}$, the required number of scalar fields is order of $10^{2}$, and hence compatible with various low energy effective models. If it is the case, then this implies that at the Planck scale, there were order one hundred number of scalar fields that have Planck scale order masses, and hence it is natural to apply the no-boundary proposal.


\section{\label{sec:dis}Discussion}

In conclusion, we summarize our results and comment for other opinions.

\subsection{\label{sec:dis1}Summary of this paper}

In this paper, we supposed that a good inflationary model should satisfy two properties:
\begin{itemize}
\item[1.] Consistent with observational constraint.
\item[2.] Can satisfy the typicality condition of the no-boundary proposal.
\end{itemize}
To satisfy the typicality condition, there are three possibilities. (P1) If the first inflation occurred in the high energy scale and there was a series of inflation, then it can be a good hypothesis and it can remain some imprints in the observational universe. (P2) If the potential is finely tuned or the field space is unbounded, then it can be a good hypothesis. (P3) If there are sufficient number of fields that can contribute to inflation (perhaps, in the Planck scale, order of one hundred), then it can be a good hypothesis. For all cases, we have some freedom to make the models consistent with observational constraints, e.g., by introducing a curvaton.

We may further impose the third condition for a good hypothesis:
\begin{itemize}
\item[3.] Less fine-tunings or unnatural assumptions.
\end{itemize}
We illustrate potential strong points and weak points for each possibilities in terms of third condition:
\begin{itemize}
\item[P1.] If the first inflation occurred in the high energy scale and there was a series of inflation,
\begin{itemize}
\item[a.] Strong points: The pre-inflation can naturally resolve some doubts on inflation.
\item[b.] Weak points: We need to adjust the smooth transition between each inflationary stages. If the secondary inflation gives more than $\mathcal{N}_{*}$ $e$-foldings, then there may be no observational consequences.
\end{itemize}
\item[P2.] If the potential is finely tuned or unbounded,
\begin{itemize}
\item[a.] Strong points: It is directly consistent with observational constraints without any other inputs.
\item[b.] Weak points: We need too strong fine-tunings. For unbounded inflation models, we need to be sure that the model is still valid for very large field values.
\end{itemize}
\item[P3.] If there are sufficient number of fields that contribute inflation,
\begin{itemize}
\item[a.] Strong points: If we introduce a curvaton, then this model can be consistent with observations and compatible with the no-boundary proposal without fine-tunings of potentials.
\item[b.] Weak points: We may need too many number of fields.
\end{itemize}
\end{itemize}

We do not mean that these weak points are fatal. If one can naturally explain the weak points, then this can be a good inflation model that satisfy three requirements. In our subjective points of view, the followings can be a good inflation model: regarding (P1), some moduli/dilaton potentials for stabilizations can naturally require and realize the pre-inflationary model; regarding (P2), if there is a justification (perhaps, from string theory) that we can justify the entire long flat direction of the potential (in Einstein frame), then this model has no problem (e.g., models from conformal inflation can justify such kind of finely tuned models \cite{Kallosh:2013hoa}); regarding (P3), if string theory naturally allow a lot of axions, and the same arguments in this paper can be applied for axion $N$-flation, then this is definitely a good model.

\subsection{General comments}

Finally, we give some comments on inflation models and general opinions. Of course, we cannot rule out a model or an idea completely, since there are always freedom to modify the original hypothesis. However, we can definitely say for some \textit{naive} traditional models or ideas, so that they are not quite natural in theoretical grounds. We illustrate the examples.
\begin{enumerate}
\item Naive slow-roll single field inflation models cannot be a good hypothesis in terms of the no-boundary proposal, since the energy scale of inflation is very small, $V_{0} \ll 1$, and hence any hierarchy of the Euclidean action will be amplified and near the cutoff region (with small $e$-foldings) will be highly preferred.

Therefore, a slow-roll single field inflation model needs an explanation: for example, it should require a pre-inflationary background as (P1) and the pre-inflation gives a proper initial condition for the secondary inflation.

\item Naive volume weighting on the no-boundary wave function \cite{Hartle:2007gi,Hartle:2008ng,Hawking:2002af} may not compatible with the observation. In terms of the number of $e$-foldings, in general, $\mathcal{N}_{\mathrm{h}} \ll \mathcal{N}_{\mathrm{c}}$, and hence, the volume-weighting prefers chaotic inflation.

Therefore, if one wants to make the volume weighting more consistent with the observations, we need more explanations, e.g., there is a limitation of the field value in the chaotic inflation part in left of Figure~\ref{fig:potential}. If we should have to do this, then the volume weighting idea becomes less attractive. We think that (P3) is a better and natural hypothesis.

\item Naive doubts on inflation cannot survive, since we can build compatible models, although they may not be a naive single field inflation model. In addition, naive doubts on the no-boundary proposal also cannot survive, by the same reasons.
\end{enumerate}

One good news is that, if there was a pre-inflationary era, then it can remain some imprints on the density perturbations, although the details remained for a future work. If it is the case, this opens the possibility that we may see the estimations of quantum cosmology from the observational data. Of course, in the present paper, we only give the guideline for model-builders; the details of the observational consequences relies on the choice of detailed model.

We need further generalizations. For example, in this paper, we only considered that all matter fields are minimally coupled each other. If there are complicated couplings or there are modified terms on the gravity sector (recently, in massive gravity, see \cite{Sasaki:2013nka}), these may affect new things. Generalization to $N$-flation with axion type fields and the no-boundary proposal is also an interesting topic. We remain these issues for interesting future work.

\section*{Acknowledgment}
We thank to Misao Sasaki and Wan-il Park for advices from the beginning of this project. DY and DH are supported by the National Research Foundation of Korea(NRF) funded by the Korea government(MEST, 2005-0049409) through the Center for Quantum Spacetime(CQUeST) of Sogang University. DY is supported by the JSPS Grant-in-Aid for Scientific Research (A) No.~21244033.

\end{document}